\documentclass[aps,prd,final,groupedaddress,%
               showpacs,showkeys]{revtex4}       

\usepackage[latin1]{inputenc}                    
\usepackage{graphicx}                            
\usepackage{latexsym}                            
\usepackage{amsfonts}                            
\usepackage{amssymb}                             
\usepackage{amsmath}                             
\usepackage[mathscr]{eucal}                      
\usepackage{dcolumn}                             
\usepackage{theorem}                             
\usepackage{hyperref}                            

\graphicspath{{.}{Graphics/}}                    
\newcommand{\imag}{\mbox{i}}                     
\newcommand{\msbar}{$\overline{\mbox{\rm MS}}$}  
\newcommand{\ts}{\tau}                           

\hypersetup{%
     debug=false,                                
     a4paper=true,                               
     dvipdf,                                     
     pdfpagemode={UseOutline},                   
     pdftitle={Moments of Nucleon Generalized Parton Distributions 
               in Lattice QCD},                  
     pdfauthor={LHPC/MIT collaboration},         
     pdfsubject={Phys Rev D publication},        
     pdfcreator={Wolfram Schroers},              
     pdfproducer={Wolfram Schroers},             
     pdfkeywords={12.38.Gc,13.60.Fz,Generalized parton distribution,
                  lattice matrix elements}       
     }

\theoremstyle{plain}
\theorembodyfont{\slshape}
\theoremheaderfont{\scshape}

\begin{document}

\preprint{MIT-3353} \title[Moments of Nucleon GPDs in Lattice
QCD]{Moments of Nucleon Generalized Parton Distributions in Lattice
  QCD}
\author{LHPC and SESAM Collaborations}
\affiliation{\quad}
\author{Ph.~H{\"a}gler}
\email{haegler@lns.mit.edu}
\author{J.~W.~Negele}
\email{negele@mitlns.mit.edu}
\author{D.~B.~Renner}
\email{dru@mit.edu}
\author{W.~Schroers}
\email{Wolfram.Schroers@Field-theory.org}
\homepage{http://www.Field-theory.org}
\affiliation{Center for Theoretical Physics\\Laboratory for Nuclear
  Science and Department of Physics\\ Massachusetts Institute of
  Technology\\Cambridge, MA 02139}
\author{Th.~Lippert}
\email{lippert@theorie.physik.uni-wuppertal.de}
\author{K. Schilling}
\email{schillin@theorie.physik.uni-wuppertal.de}
\affiliation{Department of Physics, University of Wuppertal, D-42097
  Wuppertal, Germany}

\begin{abstract}
  Calculation of moments of generalized parton distributions in
  lattice QCD requires more powerful techniques than those previously
  used to calculate moments of structure functions. Hence, we present
  a novel approach that exploits the full information content from a
  given lattice configuration by measuring an overdetermined set of
  lattice observables to provide maximal statistical constraints on
  the generalized form factors at a given virtuality, $t$. In an
  exploratory investigation using unquenched QCD configurations at
  intermediate sea quark masses, we demonstrate that our new technique
  is superior to conventional methods and leads to reliable numerical
  signals for the $n=2$ flavor singlet generalized form factors up to
  3 GeV$^2$.  The contribution from connected diagrams in the flavor
  singlet sector to the total quark angular momentum is measured to an
  accuracy of the order of one percent.
\end{abstract}



\maketitle

%
%

\section{\label{sec:Introduction}Introduction}
Light-cone correlation functions play a special role in the
experimental exploration of the quark and gluon structure of hadrons.
Asymptotic freedom allows quantitative separation of the reaction
mechanism from the structure of the probed hadron at high energy, so
that spin-independent scattering experiments unambiguously measure
matrix elements of the light-cone operator
\begin{equation}
  \label{eq:light-cone-op}
  {\cal O}(x) = \int \frac{d \lambda}{4 \pi} e^{i \lambda x} \bar
  \psi (-\frac{\lambda}{2}n)
  \not n {\cal P}e^{-ig \int_{-\lambda / 2}^{\lambda / 2} d \alpha \, n
    \cdot A(\alpha n)}
  \psi(\frac{\lambda}{2} n)\,,
\end{equation}
where $n$ is a light-cone vector, and ${\cal P}$ denotes a
path-ordering of the gauge fields in the exponential. Since these
matrix elements are singled out by their experimental accessibility,
it is essential to use all our tools of analytical methods and lattice
field theory to explore and understand them as fully as possible.

Diagonal nucleon matrix elements, $q(x) = \langle P |{\cal O}(x) | P
\rangle$, measure the familiar quark distribution $q(x)$ specifying
the probability of finding a quark carrying a fraction $x$ of the
nucleon's momentum in the light cone frame.  Although light cone
correlation functions cannot be calculated directly in lattice QCD,
expansion of ${\cal O}(x) $ generates the tower of twist-two
operators,
\begin{equation}
  \label{eq:gen-loc-curr}
  {\cal O}_q^{\lbrace\mu_1\mu_2\dots\mu_n\rbrace} = \overline{\psi}_q
  \gamma^{\lbrace\mu_1} \imag\overleftrightarrow{D}^{\mu_2} \dots
  \imag\overleftrightarrow{D}^{\mu_n\rbrace} \psi_q\,,
\end{equation}
with $\overleftrightarrow{D}=\frac{1}{2}\left( \overrightarrow{D} -
  \overleftarrow{D} \right)$, and curly braces $\lbrace
\mu_1\dots\mu_n\rbrace$ mean symmetrization of indices and subtraction
of traces. The diagonal matrix elements $ \langle P | {\cal
  O}_q^{\lbrace\mu_1\mu_2\dots\mu_n\rbrace} | P \rangle$ can be
calculated on the lattice and specify the $(n-1)^{th}$ moments $\int
dx\, x^{n-1} q(x) $. Note that expressions analogous to
Eqns.~(\ref{eq:light-cone-op}) and (\ref{eq:gen-loc-curr}) for
spin-dependent observables differ only in their gamma matrix
structure, but will not be considered in the present work.

Generalized parton distributions (GPDs), as introduced in
\cite{Muller:1994fv,Ji:1997ek,Radyushkin:1997ki}, correspond to
non-diagonal matrix elements $\langle P' |{\cal O}(x) | P \rangle$.
When expressed in terms of the relevant Lorentz invariants, $\langle
P' |{\cal O}(x) | P \rangle$ is specified in terms of two generalized
parton distributions, $H(x, \xi, t) $ and $E(x, \xi, t) $, depending
on three kinematical variables. In terms of the four-momentum transfer
$\Delta = P' - P$, the invariant momentum transfer squared is $ t=
\Delta^2$, the skewedness is $\xi = -n \cdot \Delta /2$, and $x$
denotes the momentum fraction.  Since the dependence of the GPDs,
$H(x, \xi, t) $ and $E(x, \xi, t) $, on three kinematical variables
renders their physical interpretation more difficult than ordinary
parton distributions, it is useful to recall several important
physical properties. In the forward limit, i.e.~$\xi$, $t \to 0$, we
recover the forward parton distribution function as $H(x,0,0) = q(x)$.
In what is sometimes referred to as the local limit, integrating over
the momentum fraction, $x$, yields the familiar electromagnetic form
factors, $ \int dx\; H(x, \xi, t) = F_1(t)$ and $ \int dx\; E(x, \xi,
t) = F_2(t)$.  The first moment of the sum of $H$ and $E$ yields the
total quark angular momentum $\frac{1}{2} \int dx\; x [H(x, \xi, t) +
E(x, \xi, t) ] = J_q$. We note that both these results are independent
of $\xi$. And finally, at skewedness $\xi = 0$, the $ t$ dependence
specifies the transverse Fourier transform with respect to the impact
parameter of the light cone wave function~\cite{Burkardt:2000za}. For
a possible interpretation of GPDs with non-zero $\xi$, see
Refs.~\cite{Ralston:2001xs,Diehl:2002he}.

On the lattice, instead of matrix elements of the light cone operator,
Eq.~(\ref{eq:light-cone-op}), one again calculates non-diagonal matrix
elements of the local operators, Eq.~(\ref{eq:gen-loc-curr}), yielding
moments of the generalized parton distributions.  Following the
notation of Ref.~\cite{Ji:1997ek}, the non-diagonal matrix element
$\langle P' |{\cal O}^{\lbrace\mu_1\dots\mu_n\rbrace} | P \rangle$ may
be expressed in terms of a set of generalized form factors (GFFs)
$A_{ni} (t) $, $B_{ni} (t) $, and $C_{n} (t) $. The form factors
$A_{ni} (t) $ and $C_{n} (t) $ multiplied by powers of $\xi^2$ yield
the moment $H_n(\xi,t) = \int dx\; x^{n-1} H(x,\xi,t)$ and the form
factors $B_{ni}(t) $ and $C_{n}(t) $ multiplied by powers of $\xi^2$
yield the moment $E_n(\xi,t) = \int dx\; x^{n-1} E(x,\xi,t)$.

The lowest three moments considered in this work are
\begin{eqnarray}
  \label{eq:explicit}
  \langle P' | {\cal O}^{\mu_1} | P \rangle &=&
  \langle \! \langle \gamma^{\mu_1 }\rangle \! \rangle A_{10}(t) 
  \nonumber \\
  & & \quad + \frac{\imag}{2 m} \langle \! \langle \sigma^{\mu_1 
    \alpha} \rangle \! \rangle
  \Delta_{\alpha} B_{1
    0}(t)\,, \nonumber \\ [.5cm]
  \langle P' | {\cal O}^{\lbrace \mu_1 \mu_2\rbrace} | P \rangle &=&
  \bar P^{\lbrace\mu_1}\langle \! \langle
  \gamma^{\mu_2\rbrace}\rangle  \! \rangle
  A_{20}(t) \nonumber \\
  & & \quad + \frac{\imag}{2 m} \bar P^{\lbrace\mu_1} \langle \! \langle
  \sigma^{\mu_2\rbrace\alpha}\rangle \! \rangle \Delta_{\alpha} B_{2 0}(t)
  \nonumber \\
  & & \quad +\frac{1}{m}\Delta^{\{ \mu_1}   \Delta^{ \mu_2 \} }
  \langle \! \langle 1 \rangle \! \rangle C_{2}(t)\,, \nonumber \\[.5cm]
  \langle P' | {\cal O}^{\lbrace\mu_1 \mu_2 \mu_3\rbrace} | P \rangle
  &=& \bar P^{\lbrace\mu_1}\bar P^{\mu_2} \langle \! \langle
  \gamma^{\mu_3\rbrace}
  \rangle \! \rangle A_{30}(t) \nonumber \\
  & & \quad + \frac{\imag}{2 m} \bar P^{\lbrace \mu_1}\bar P^{\mu_2}
  \langle \! \langle \sigma^{\mu_3\rbrace\alpha} \rangle \! \rangle
  \Delta_{\alpha} B_{3 0}(t) \nonumber
  \\
  & & \quad + \Delta^{\lbrace \mu_1}\Delta^{\mu_2} \langle \! \langle
  \gamma^{\mu_3\rbrace}\rangle \! \rangle A_{32}(t) \nonumber \\
  & & \quad +\frac{\imag}{2 m} \Delta^{\lbrace\mu_1}\Delta^{\mu_2}
  \langle \! \langle \sigma^{\mu_3\rbrace\alpha}\rangle \! \rangle
  \Delta_{\alpha} B_{3 2}(t)\,,
\end{eqnarray}
where $\bar P_{\mu} = (P_{\mu} + P'_{\mu})/2 $ and $ \langle \!
\langle \Gamma \rangle \! \rangle = \bar U(P') \Gamma U(P)$.  The
GFFs, $A_{ni} (t) $, $B_{ni} (t) $, and $C_{n} (t) $, specify all the
information about spin-independent generalized parton distributions
that is known to be accessible on the lattice.  The limits for $H$ and
$E$ discussed above may be re-expressed in terms of the generalized
form factors. The limit $t \to 0$ of $A_{n0}$ is the familiar parton
distribution moment, $A_{n0}(0) = \int dx\; x^{n-1} q(x)$.  The
electromagnetic form factors are given by $A_{10}(t) = F_1(t)$ and $
B_{10}(t) = F_2(t)$ for the appropriate flavor combination.  Finally,
the total quark angular momentum is given by the sum of $A_{20}+
B_{20}$ as $ t \to 0$, $ J_q = \frac{1}{2} [ A_{20} (0) + B_{20}(0) ]
$.

In the context of this brief review, we may now consider the
challenges and opportunities in calculating moments of generalized
parton distributions on the lattice, and compare them with the
analogous issues for ordinary parton distributions. Experimentally,
three decades of deep inelastic scattering experiments have provided
impressive phenomenological determinations of parton distributions as
a function of momentum fraction \cite{Lai:1997mg,Gluck:1998xa,%
  Martin:2001es}. Hence, for parton distributions, the key issues are
developing the lattice technology to the point of attaining
quantitative agreement with moments of experimental parton
distributions and using the lattice as a tool to obtain insight into
how these distributions arise from QCD\@. The present status is that
computational limitations restrict unquenched QCD calculations to the
heavy quark regime in which the pion mass is heavier than roughly
$500\,$MeV and naive linear extrapolation to the physical pion mass
yields serious disagreement with experiment. For example, linear
extrapolation of the quark momentum fraction exceeds experiment by the
order of fifty percent~\cite{Gockeler:1996wg,Best:1997ab,%
  Gockeler:1997jk,Gockeler:2001us,Dolgov:2002zm}.  Although there are
strong indications that physical extrapolation to the chiral limit may
introduce corrections of the required magnitude \cite{Detmold:2001jb},
there is presently no quantitative theory for the intermediate mass
regime and we must exploit the emerging generation of computers to
perform the requisite calculations sufficiently close to the chiral
regime.

For generalized parton distributions, the situation is quite
different.  Since GPDs depend on three variables and experimental
quantities involve convolutions, there is no prospect of measuring the
full dependence on $x$, $\xi$, and $t$.  Without additional input
arising from first principles, extraction of GPDs from experiments
like deeply virtual Compton scattering will necessarily be
contaminated by uncontrolled assumptions.  Hence, once computer power
is sufficient to obtain quantitative agreement with moments of parton
distributions, lattice calculations of moments of GPDs will become an
essential tool to be used in conjunction with experiment to extract
and understand the full dependence on $x$, $\xi$, and $t$. It is thus
imperative to develop techniques to calculate these moments.

In addition, theorists may also obtain insight into how QCD works by
studying the dependence of hadron structure on the quark mass.  This
study can begin immediately, addressing the behavior of hadrons in a
world where the pion weighs more than $500\,$MeV. This heavy pion
world is much closer to the non-relativistic quark model, and as we
eventually lower the pion mass, we will learn how QCD evolves from the
world of heavy quarks to the physical world of light quarks. Even in
the heavy pion world, one can test contemporary assumptions, such as
factorization of the $t$ dependence~\cite{Belitsky:2001ns}.

On the computational side, it is essential to confront the additional
challenges that arise for GPDs relative to the ordinary parton
distributions. Already for forward parton distributions, the tower of
operators, Eq.~(\ref{eq:gen-loc-curr}), involves operators that become
increasingly subject to statistical noise as one progresses to higher
and higher derivatives. For GPDs, however, we compound the noise of
these operators with the additional noise from the finite momentum
transfer, $\Delta$.  Hence, in this work, we address the problem of
imposing the maximal statistical constraints a lattice calculation can
provide on the form factors $A_{ni}$, $B_{ni}$, and $C_{n}$ appearing
in Eq.~(\ref{eq:explicit}).  We note that we have at our disposal the
choice of several alternative representations of the hypercubic group
on the lattice corresponding to the same continuum operator ${\cal
  O}^{\lbrace\mu_1\mu_2\dots\mu_n\rbrace}$ as well as the choice of
different kinematic variables corresponding to the same $t$.  We will
therefore use this freedom to construct an over-determined set of
lattice observables corresponding to the continuum expressions in
Eq.~(\ref{eq:explicit}), and thereby significantly improve the
measurement of the form factors.  In section II, we will describe the
details of the method. Section III will compare the results of our
method with a conventional analysis and present results for the the
angular momentum carried by quarks, $J_q$. The conclusions will be
presented in the final section and an Appendix presents the necessary
detailed expressions for matrix elements in terms of generalized form
factors.

%
%

\section{\label{sec:lattice-techniques}Lattice Calculation of
  Generalized Form Factors}
We will calculate the matrix elements of $ \langle P' | {\cal
  O}^{\lbrace\mu_1\mu_2\dots\mu_n\rbrace}_q | P \rangle$ to extract
the generalized form factors $A_{ni}^q(t)$, $B_{ni}^q(t)$, and
$C_{n}^q(t)$ where, when relevant, we append a quark flavor label $q$.
Since the calculation of disconnected diagrams raises yet another
level of complexity, in this present work we will restrict our
attention primarily to the flavor non-singlet combination $u-d$, for
which these diagrams cancel.

In the usual way, we calculate the non-diagonal matrix elements by the
following ratio of three- and two-point functions,
\begin{equation}
  \label{eq:ratio}
  R_{\cal O}(\ts,P',P) = \frac{C_{\cal O}^{\text{3pt}}(\ts,P^{\prime
    },P)}
  {C^{\text{2pt}}(\ts_{\text{snk}},P^{\prime })}\left[
    \frac{C^{\text{2pt}}(\ts_{\text{snk}}-\ts+\ts_{\text{src}},P)
      \;C^{\text{2pt}}(\ts,P')\;C^{\text{2pt} }(\ts_{\text{snk}},P')}%
    {C^{\text{2pt}}(\ts_{\text{snk}}-\ts+\ts_{\text{src}},P')%
      \;C^{\text{2pt}}(\ts,P)\;C^{\text{2pt}}(\ts_{\text{snk}},P)}
  \right]^{1/2}\,.
\end{equation}
The factors relating this ratio to the physical continuum matrix
element are given in Eq.~(\ref{eq:RatioM}). The correlation functions
are given by
\begin{eqnarray}
  \label{eq:n-point-def}
  C^{\text{2pt}}(\ts,P) &=& \sum_{j,k}\left(
    \Gamma_{\text{unpol}}\right)_{jk}\left\langle \Omega\right|
  N_{k}(\ts,P)\overline{N}_{j}(\ts_{\text{src}},P)\left|
    \Omega\right\rangle\,,
  \nonumber\\
  C_{\cal O}^{\text{3pt}}(\ts,P',P) &=& \sum_{j,k}\left(
    \Gamma_{\text{pol}}\right)_{jk}\left\langle \Omega\right|
  N_{k}(\ts_{\text{snk}},P'){\cal O}(\ts)
  \overline{N}_{j}(\ts_{\text{src}},P)\left| \Omega\right\rangle\,,
\end{eqnarray}
where $\vert\Omega\rangle$ denotes the QCD vacuum state.  The nucleon
source, $\bar{N}(\ts,P)$, and sink, $N(\ts,P)$, create and annihilate
states with the quantum numbers of the nucleon, and to maximize the
overlap with the ground state, we used the smeared sources defined in
Ref.~\cite{Dolgov:2002zm}. The source is located at timeslice
$\ts_{\text{src}}$, the operator ${\cal O}$ is inserted at timeslice
$\ts$, and the sink is positioned at timeslice $\ts_{\text{snk}}$.
Explicit expressions for the polarized and unpolarized projectors
$\Gamma _{\text{pol/unpol}}$ are given in
Appendix~\ref{sec:matr-elem-terms},
Eqs.~(\ref{eq:LHPC_proj},\ref{eq:QCDSF_proj}).

Inserting a complete set of states into Eq.~(\ref{eq:n-point-def}) and
using the time evolution operator yields
\begin{eqnarray}
  \label{eq:expand-2pt}
  C^{\text{2pt}}(\ts,P) &=& \sum_l
  e^{-E_l(P)(\ts-\ts_{\text{src}})} \mbox{Tr}\left(
    \Gamma_{\text{unpol}} \langle\Omega\vert N(\ts,P)\vert l\rangle
    \langle
    l\vert \bar{N}(\ts_{\text{src}},P)\vert\Omega\rangle\right)
  \nonumber
  \\
  &=& e^{-E_0(P)(\ts-\ts_{\text{src}})} \frac{\left(
      Z(P)\overline{Z}(P)\right)^{1/2}}{E_0(P)} \;\mbox{Tr}\left(
    \Gamma_{\text{unpol}} U(P)\overline{U}(P)\right) \nonumber \\
  && \qquad + \mbox{higher states} \,, \\
  \label{eq:expand-3pt}
  C_{\cal O}^{\text{3pt}}(\ts,P',P) &=&
  \sum_{k,l}\mbox{Tr}\Biggl\lbrace
  \Gamma_{\text{pol}}\langle\Omega\vert N(\ts_{\text{snk}},P') \vert
    k\rangle\langle k\vert e^{-E_k(P')\left(\ts_{\text{snk}}-\ts
      \right)}  \nonumber \\
  && \quad\times{\cal O} \vert l\rangle\langle l\vert
    e^{-E_l(P)\left(\ts-\ts_{\text{src}}\right)} 
  \vert \overline{N}\left(\ts_{\text{src}},P\right)\vert
  \Omega\rangle\Biggr\rbrace \nonumber \\
  &=&
  e^{-E_0(P)(\ts-\ts_{\text{src}})-E_0(P')(\ts_{\text{snk}}-\ts)}
  \frac{\left(
      Z(P)\overline{Z}(P')\right)^{1/2}}{E_0(P')E_0(P)}\mbox{Tr}\left(
    \Gamma_{\text{pol}}U(P')\overline{U}(P)\right) \nonumber\\
  && \quad\times \left\langle P'\right| {\cal
    O}(\ts)\left| P\right\rangle + \mbox{higher states} \,.
\end{eqnarray}
The contributions from higher states in Eqns.~(\ref{eq:expand-2pt})
and (\ref{eq:expand-3pt}) are suppressed by exponential prefactors
when $\ts_{\text{snk}}-\ts$ and $\ts-\ts_{\text{src}}$ are
significantly greater than the inverse of the excitation energy of the
first excited state.

The ratio, Eq.~(\ref{eq:ratio}), is constructed to exactly cancel all
exponential and wave-function overlap factors. The two-point functions
$C^{\text{2pt}}(\ts_{\text{snk}},P^\prime)$ and
$C^{\text{2pt}}(\ts_{\text{snk}},P)$ decay exponentially for the full
Euclidean distance between the source and sink and are thus
particularly subject to statistical noise with finite statistics.  In
the worst case, they may even become negative, and these cases are
excluded from the present work.  We note that other possibilities
besides Eq. 4 may be used to cancel the exponential and overlap
factors, and this freedom will be explored in a subsequent work.

For sufficiently large time separations the ratio $R(\ts,P',P)$ will
exhibit a plateau yielding the desired lattice matrix element, and the
plateau value, $\bar{R}(P',P)$, is obtained by averaging over an
appropriate range of timeslices, $\ts_{\text{min}}$ to
$\ts_{\mbox{\tiny max}}$,
\begin{equation}
  \bar{R}_{\lbrace\mu_1\mu_2\dots\mu_n\rbrace}(P',P) =
  \frac{1}{\ts_{\text{max}}- \ts_{\text{min}}}
  \sum_{\ts=\ts_{\mbox{\tiny
        min}}}^{\ts_{\text{max}}}R_{\lbrace\mu_1\mu_2\dots\mu_n\rbrace} 
  (\ts,P',P)\,.
\end{equation}
To convert our lattice calculations to the continuum \msbar-scheme, we
use 1-loop perturbative matching at the scale $\mu^2=4\text{GeV}^2$
\begin{equation}
  \label{eq:cont-me-def}
  {\cal O}^{\overline{\text{MS}}}_i(\mu) = \sum_j Z_{ij}(\mu,a) {\cal
    O}^{\text{lat}}_j(a) \,,
\end{equation}
so that the lattice matrix element is related to the continuum matrix
element by
\begin{equation}
  \label{eq:cont-compute}
  \left\langle P'\right| {\cal O}^{\overline{\text{MS}}}_i 
  \left|P\right\rangle = \sqrt{E(P^\prime)E(P)} \sum_j Z_{ij}
  \overline{R}_j\,.
\end{equation}
Note that the renormalization constant, $Z_{\cal O}(\mu,a)$, depends
only on the operator ${\cal O}$, but not on the external states, so
that Eq.~(\ref{eq:cont-compute}) is valid for any external momenta,
$P'$ and $P$.

Finally, we write the Euclidean continuum relation between the renormalized
matrix element of the generalized current 
$\langle P'\vert{\cal O}^q_{\lbrace \mu_1\mu_2\dots\mu_n\rbrace}\vert P\rangle$ and the
desired generalized form factors $A^q_{ni}(t), B^q_{nj}(t)$, and $
C^q_{n}(t)$ in the following abbreviated notation:
\begin{equation}
  \label{eq:r-mink}
  \left\langle P'\right| {\cal O}^q_{\lbrace\mu_1\mu_2\dots\mu_n\rbrace}\left|
    P\right\rangle = \sum_i a_i A^q_{ni} + \sum_j b_j B^q_{nj}
  + c C^q_{n}\,.
\end{equation}
Full expressions for the kinematic factors $\lbrace a_i, b_j,
c\rbrace$ are given in Appendix~\ref{sec:matr-elem-terms} for $n=1,2,$
and $3$, Eq.~(\ref{eq:parametrization}).

For a given $n$, we may evaluate the $N_n^{GFF}$ generalized form
factors $\lbrace A^q_{ni}(t), B^q_{nj}(t), C^q_{n}(t)\rbrace$ as
follows. We select $N \ge N_n^{GFF}$ sets of operators ${\cal
  O}^q_{\lbrace\mu_1\mu_2\dots\mu_n\rbrace}$ and momenta $\{P', P\}$ such that
Eq.~(\ref{eq:r-mink}) specifies $N_n^{GFF}$ linearly independent
combinations of the form factors.  Lattice matrix elements for these
operators and momenta are calculated from the the ratios
$R(\ts,P',P)$, in Eq.~(\ref{eq:ratio}) and matched to continuum
operators via Eq.~(\ref{eq:cont-me-def}). If $N = N_n^{GFF}$, the GFFs
are calculated by inverting Eq.~(\ref{eq:r-mink}), and if $N >
N_n^{GFF}$, they are calculated by a least-squares fit to the
overdetermined system Eq.~(\ref{eq:r-mink}).

Note, that in contrast to the case of forward parton distributions,
where the moments correspond to a single number (denoted $v_{n}$ or
$\langle x^{n-1} \rangle$), or electromagnetic form factors, where
there are two form factors ($F_1$ and $ F_2$), we have the
complication of $N_n^{GFF}$ unknown generalized form factors to be
determined.  We therefore now discuss the strategy for selecting an
appropriate set of operators and momenta for this task.

\subsection{\label{sec:pract-cons}Practical considerations}
One practical concern is the numerical noise associated with momentum
projection.  The three-point function is subject to noise from the
projection of the sink onto momentum $P'$ and of the operator onto
momentum transfer $\Delta $.  In addition to being subject to the sink
momentum projection, the two-point functions appearing in the lattice
ratio also collect noise from the projection onto the source momentum
$P$.  Ideally, for each invariant momentum transfer $t$, one would
like to select $P$ and $P'$ such as to minimize these errors, but each
distinct choice of sink momentum $P'$ requires an expensive
calculation of a new set of backward propagators, whereas changing $
\Delta $ for fixed $P'$ requires no new propagators.  Hence, denoting
the lowest momentum attained on a lattice with $L_s$ lattice points in
a spatial dimension as $p_l=2\pi/L_s$.  As a practical matter, we have
used two sink momenta, $\vec{P}\,' = (0,0,0) $ and $\vec{P}\,' =
(-p_l,0,0)$.  With these sinks, we generate a substantial set of
momenta listed in Appendix~\ref{sec:matr-elem-terms},
Tab.~\ref{tab:all-latt-moms}, by using a range of values of $\Delta$.
However there are two special virtualities of note, the virtuality,
$t_{\text{vlow}}$ and the Breit frame virtuality $t_{\text{Breit}}$
given by the following momentum combinations, or rotations thereof:
\begin{eqnarray}
  \label{eq:vlow-def}
  \vec{P}\,' = (0,0,0)\,, \vec{P} = (p_l,0,0) &\Rightarrow&
  t_{\text{vlow}} = \left(m-\sqrt{p_l^2+m^2}\right)^2 - p_l^2\,, \\
  \label{eq:breit-def}
  \vec{P}\,' = (-p_l,0,0)\,, \vec{P} = -\vec{P}\,' &\Rightarrow&
  t_{\text{Breit}} = 4p_l^2\,,
\end{eqnarray}
The virtuality $t_{\text{vlow}}$ is the lowest virtuality that can be
placed on the lattice for the set of external momenta we use, for
which no spatial momentum exceeds $p_{l}$, and would be expected to
have minimal projection error.  Since the time component does not
reduce the virtuality in the Breit frame, the Breit frame provides the
optimal means of providing a large virtuality with minimal momentum
projection error.  Hence, including both $t_{\text{vlow}}$ and
$t_{\text{Breit}}$ in our full set of momentum selections ensures the
presence of measurements at both ends of our $t$-range that have the
minimal possible momentum-induced statistical error. In addition to
these considerations, for certain different virtualities $t$, there
are more available momentum combinations and thus more constraints
than in the Breit frame. Hence, we obtain comparably small errors for
most of the other virtualities as well.

Another practical issue in measuring generalized form factors is the
presence of powers of the momentum transfer $\Delta$ in the
kinematical factors in Eq.~(\ref{eq:r-mink}).  The four-momentum
$\Delta$ is --- compared to the time component of external momenta in
the cases we consider --- a small number, so its presence in
kinematical factors amplifies the effect of statistical errors in
lattice matrix elements on measurements of the associated form
factors.  Consistent with the pattern observed in
Eq.~\ref{eq:explicit}, in the general case the factors $a_i$
multiplying $A^{q}_{n i}(t)$ contain $i=\{0, 2, \cdots
2[\frac{n-1}{2}] \}$ factors of $\Delta$, the factors $b_j$
multiplying $B^{q}_{n j}(t)$ contain $j=\{1, 3, \cdots
1+2[\frac{n-1}{2}]\}$ factors of $\Delta$, and the factor $c$, which
is non-vanishing only for even $n$, has $n$ factors of $\Delta$.
Therefore, we expect $A^{q}_{n0}(t)$ to be the quantity that can be
extracted most accurately from lattice calculations, $B^{q}_{n0}(t)$
should have slightly larger errors, $C^{q}_{n}(t)$ is the worst
determined when it is non-vanishing, and the remaining GFFs,
$A^{q}_{ni}(t)$ and $B^{q}_{nj}(t)$, should lie somewhere in between
these extremes.

\subsection{\label{sec:overd-set-latt}Overdetermined set
  of lattice observables}
To extract generalized form factors from lattice calculations, we now
consider a fixed value of the virtuality, $t$, that can be achieved on
the lattice, and abbreviate Eqns.~(\ref{eq:cont-compute}) and
(\ref{eq:r-mink}) in the following schematic form:
\begin{eqnarray}
  \label{eq:gff-soe1}
  \langle {\cal O}_i^{cont}\rangle &=& \sum_j a_{ij} {\cal F}_j\,,
  \nonumber \\
  \langle {\cal O}_i^{cont}\rangle &=& \sqrt{E' E} \sum_j Z_{ij}
  \overline{R}_j\,,
\end{eqnarray}
where we denote the generalized form factors generically by ${\cal
  F}_j(t)$ and $j$ runs over all the form factor labels for the $n$
under consideration. Thus, for n = 1, $ {\cal F}_1(t)=A^q_{10}(t)$ and
$ {\cal F}_2(t)=A^q_{20}(t)$. Note that the familiar spin non-flip and
spin-flip electromagnetic form factors are defined by these flavor
form factors weighted with the electric charge, i.e.~$F_1(t)=2/3
A^{\text u}_{10}(t) - 1/3 A^{\text d}_{10}(t)$, and $F_2(t)=2/3
B^{\text u}_{10}(t) - 1/3 B^{\text d}_{10}(t)$. Similarly, for $n=2$,
corresponding to the operator ${\cal O}_{\lbrace\mu\nu\rbrace}$, the
${\cal F}_i(t)$ in Eq.~(\ref{eq:gff-soe1}) denote ${\cal
  F}_1(t)=A^q_{20}(t)$, ${\cal F}_2(t)=B^q_{20}(t)$, and ${\cal
  F}_3(t)=C^q_{2}(t)$.

Eliminating $\langle {\cal O}_i^{cont}\rangle$ from
Eqs.~(\ref{eq:gff-soe1}) yields the final form of the set of equations
we will use for our analysis:
\begin{eqnarray}
  \label{eq:gff-soe2}
  \overline{R}_i &=& \frac{1}{\sqrt{E' E}} \sum_{jk} Z^{-1}_{ij}
  a_{jk} {\cal F}_k \nonumber \\
  &\equiv& \sum_j a^{\prime}_{ij} {\cal F}_j\,.
\end{eqnarray}
In this set of equations, the label $j$ runs over the $N_n^{GFF}$
generalized form factors for the given $n$.  The label $i$ denotes a
specific choice of nucleon momenta $P$ and $P'$, and an index
combination for a lattice H(4) representation of a continuum operator
$ {\cal O}^q_{\lbrace\mu_1\mu_2\dots\mu_n\rbrace}$. We consider $N$
distinct values of this label, $i$, corresponding to the same $n$ and
$t$ and require $N > N_n^{GFF}$.  Operationally, to obtain this
collection of $N$ equations, we list all external lattice momenta,
group them together in classes of identical virtual momentum transfer,
and write down for each momentum combination the set of indices,
$\lbrace i\rbrace$, of ${\cal O}_{\lbrace i\rbrace}$ that gives rise
to non-trivial kinematic factors, $a_{ij}$. Note that sets of external
momenta in which the spatial components differ only by cubic rotations
still yield different constraints. Rotations changing components along
the spin polarization axis in general change the kinematic factors and
yield independent equations.  Rotations that change components
orthogonal to the spin polarization axis yield constraints that are
physically equivalent, but provide additional statistical constraints
in a lattice Monte Carlo calculation.
Appendix~\ref{sec:matr-elem-terms} shows the explicit momentum
combinations and the basis of diagonal index combinations we have
chosen in this work.

Equation~(\ref{eq:gff-soe2}) is an overdetermined system of linear
equations that in principle provides the maximal information on the
generalized form factors that is attainable from the lattice.  We will
compare it in Sec.~\ref{sec:comp} with the conventional technique of
using a uniquely determined system of equations. In the case of
measuring moments of parton distributions (see, e.g.
Refs.~\cite{Gockeler:1996wg,Best:1997ab,Gockeler:1997jk,%
  Gockeler:2001us,Dolgov:2002zm}) the conventional
technique corresponds to selecting a single row of
Eq.~(\ref{eq:gff-soe2}), and for electromagnetic form factors, it
corresponds to taking linear combinations or selecting kinematic
variables such that $F_1$ and $F_2$ appear separately in two
equations.

It is important to note that Eq.~(\ref{eq:gff-soe2}) contains
systematic, as well as statistical errors. The lattice action, the
discrete approximations to the operators ${\cal
  O}^q_{\lbrace\mu_1\mu_2\dots\mu_n\rbrace}$, and the nucleon
dispersion relation contain lattice artifacts, and since we use Wilson
fermions, the calculation contains errors of ${\cal O} (a)$.  In
principle these lattice artifacts can by addressed by improved
operators and extrapolation to the continuum limit, but these
corrections are beyond the scope of the present work.  In addition,
the use of 1-loop perturbative renormalization introduces systematic
errors in the analysis, which ultimately can be improved by higher
order corrections or non-perturbative renormalization.

\subsection{\label{sec:sing-value-decomp}Singular Value
  Decomposition}
We now turn to the solution of the overdetermined set of equations,
Eq.~(\ref{eq:gff-soe2}), which we write in matrix form as follows:
\begin{equation}
  \label{eq:gff-determination}
  R' = A' \cdot {\cal F}\,.
\end{equation}
Here, ${\cal F}$ is the vector of the desired GFFs, $R'$ denotes the
statistically measured lattice ratios, and $A'$ is the matrix of the
coefficients $a'_{ij}$. Changing to a standard notation, we let $n+1$
denote the number of GFFs, which we previously called $N^{GFF}$, and
continue to call the number of constraints $N$, so that $A$ is an
$N\times (n+1)$ matrix, ${\cal F}$ is a vector of length $n+1$, and
$R'$ a vector of length $N$.

We solve this overdetermined problem by minimization of the $\chi^2$
norm
\begin{equation}
  \label{eq:chisq-minn}
  \chi^2 = \sum_{i=1}^{N} \left( \frac{\sum_{j=1}^{n+1} A'_{ij} {\cal
        F}_{j} - R'_{i}}{\sigma_{i}} \right)^2\,,
\end{equation}
where $\sigma_i$ denotes the jackknife error of the lattice
measurement of $R'_{i}$. In this case the jackknife procedure should
be performed with an appropriate bin-size to eliminate autocorrelation
effects.  It is convenient to absorb the errors $\sigma_i$ by defining
$A_{ij} = A'_{ij} / \sigma_i$ and $R_i = R'_i / \sigma_i$ so that we
seek a vector ${\cal F}$ that minimizes
\begin{equation}
  \label{eq:chisq-min}
  \chi^2 = |A \cdot   {\cal F} - R |^2\,.
\end{equation}
The method of choice to solve Eq.~(\ref{eq:chisq-min}) in the presence
of singularities or near singularities is {Singular Value
  Decomposition} (SVD) \cite{Teukolsky:1992nr}. It is based on the
theorem that any $N\times (n+1)$ matrix may be decomposed as \[ A = U
\cdot \mbox{diag}\left(w_1,\dots,w_{n+1}\right)\cdot V\,, \] where $U$
is an $N\times (n+1)$ matrix, $V$ is $(n+1)\times (n+1)$, and all
$w_i$, the {singular values}\ of $A$, are non-negative.  If all the
$w_i$ are non-zero, the solution that minimizes
Eq.~(\ref{eq:chisq-min}) is given by
\begin{equation}
  \label{eq:svd-solve}
  F=V\cdot\mbox{diag}\left(1/w_1,\dots,1/w_{n+1}\right)\cdot
  \left(U^T\cdot R\right)\,,
\end{equation}
which provides the optimal vector of form factors, ${\cal F}$, we can
achieve with all the information available.

If none of the singular values is zero, the method is equivalent to
standard least-squares minimization. If one or more of the singular
values, $w_i$, are zero, the rank of the matrix $A$ is reduced by the
number of zero singular values, implying that the associated
directions in the solution space can not be explored. In the case of
zero or nearly zero singular values, we may opt to avoid exploring the
offending direction(s) and minimize Eq.~(\ref{eq:chisq-min}) in the
residual subspace by replacing the corresponding factors $1/w_i$ by
zero. In this case, of course, only the components of ${\cal F}$
corresponding to non-zero values of $w_i$ are determined.

As an example, in the forward case the singular values for all GFFs
other than $A^q_{n0}(0)$ are zero and the result is the minimization
in the subspace of $A^q_{n0}(0)$ alone. In this case, $A^q_{n0}(0$) is
still determined simultaneously from several different lattice
measurements and this procedure thereby automatically incorporates
more physical constraints than the conventional method of fitting a
single lattice observable.

The errors in ${\cal F}$ are calculated using a jackknife analysis
in the following way.
The system (\ref{eq:chisq-min}) is solved simultaneously for the full
set of lattice measurements $R$ to get the average solution vector
${\cal F}$. Then, the appropriate subsamples $R_l$ are formed from
which the subsample solutions ${\cal F}_l$ are calculated. The error
vector for the solution, $\sigma_{\cal F}$, is formed from the ${\cal
  F}_l$s by the familiar jackknife formula (see
e.g.~\cite{Schroers:2001ph} and references therein for details).

In summary, the method introduced in this section provides a way to
extract maximal information from a set of lattice calculations. It
simultaneously includes all available data in a unified manner and
corrects for correlations present in the underlying sample of gauge
field configurations.

%
%

\section{\label{sec:results}Results}
In this section, we apply our procedure to the case $n=2$ using the
operators in Eqs.~(\ref{eq:neq2diag-ops},\ref{eq:neq2offdiag-ops}) and
thoroughly analyze the results.  For this exploratory calculation, we
use SESAM unimproved Wilson configurations on a $16^3\times 32$
lattice at $\beta=5.6$ with
$\kappa_{\text{sea}}=\kappa_{\text{val}}=0.1560$.  We note that this
is the heaviest of the three quark masses used in earlier calculations
of moments of quark distributions\cite{Dolgov:2002zm}, and using the
lattice spacing $a^{-1}=2.01\,$GeV from the chirally extrapolated
nucleon mass, the pion mass $(am)_\pi=0.446(3)$ corresponds to $m_\pi
= 896(6)\,$MeV. All the computational details for calculating smeared
sources and sinks, nonrelativistic sequential sources, the difference
approximations to the operators $ {\cal
  O}^q_{\lbrace\mu_1\mu_2\dots\mu_n\rbrace}$, spin polarization
projection, and Dirichlet boundary conditions are as in
Ref.~\cite{Dolgov:2002zm}.  In this present work, we relabel the time
slices such that the source is at $\ts_{\text{src}}=1$, the sink is at
$\ts_{\text{snk}}=13$ and boundaries are at $\ts = -9$ and $\ts = 22$.

The one-loop perturbative renormalization constants for the cases we
consider in this work are taken from Ref.~\cite{Dolgov:2002zm}. Note,
that in the case of the operators ${\cal O}_{\lbrace\mu\nu\rbrace}$,
Eqs.~(\ref{eq:neq2diag-ops},\ref{eq:neq2offdiag-ops}), the
renormalization constant will depend on the index combination chosen.
The diagonal and non-diagonal index combinations belong to distinct
irreducible representations of the lattice point group and have
different renormalization factors:
$Z_{\text{diag}}^{n=2}(\mu^2=4\,\mbox{GeV}^2)=0.9768$ and
$Z_{\text{non-diag}}^{n=2}(\mu=4\,\mbox{GeV}^2)=0.9884$.

\subsection{\label{sec:comp}Comparison of Minimally Determined and 
  Overdetermined Measurements}
To explore how the method works and to compare it with the
conventional method of measuring a single operator for each
observable, we consider the case of virtuality
$t_{\text{vlow}}=-0.5925\,$GeV$^2$, which may be achieved by choosing
any of the momentum combinations displayed in Tab.~\ref{tab:low-virt}.
\begin{table}
  \caption{\label{tab:low-virt}Different combinations of external
    momenta with virtuality
    $t_{\text{vlow}}=-0.5925$GeV$^2$.}
  \begin{ruledtabular}
    \begin{tabular}{*{8}{c}}
      N                & 1           & 2           & 3           & 4
      & 5          & 6          & 7 \\ \hline
      $\vec{P}^\prime$ &$(0,0,0)$    &$(0,0,0)$    &$(0,0,0)$
      &$(0,0,0)$   &$(0,0,0)$   &$(0,0,0)$   & $(-p_l,0,0)$ \\ \hline
      $\vec{P}$        &$(-p_l,0,0)$ &$(0,-p_l,0)$ &$(0,0,-p_l)$
      &$(p_l,0,0)$ &$(0,p_l,0)$ &$(0,0,p_l)$ & $(0,0,0)$
    \end{tabular}
  \end{ruledtabular}
\end{table}

We start by considering only the combination of external momenta,
$\vec{P}\,'=(0,0,0)$, $\vec{P}=(-p_l,0,0)$, listed in the first column
in Tab.~\ref{tab:low-virt}. The system of equations,
Eq.~(\ref{eq:gff-soe2}) then contains the operators
\begin{equation}
     \label{eq:gff-oplist}
     \langle P'\vert{\cal O}^{\text{u-d}}_{\text{diag,1}}\vert
     P\rangle\,,
     \langle P'\vert{\cal O}^{\text{u-d}}_{\text{diag,2}}\vert
     P\rangle\,,
     \langle P'\vert{\cal O}^{\text{u-d}}_{\text{diag,3}}\vert
     P\rangle\,,
     \langle P'\vert{\cal O}^{\text{u-d}}_{\lbrace 10\rbrace}\vert
     P\rangle\,,
     \langle P'\vert{\cal O}^{\text{u-d}}_{\lbrace 20\rbrace}\vert
     P\rangle\,,
     \langle P'\vert{\cal O}^{\text{u-d}}_{\lbrace 21\rbrace}\vert
     P\rangle\,,
\end{equation}
with the operator index conventions given in
Appendix~\ref{sec:matr-elem-terms},
Eqs.~(\ref{eq:neq2diag-ops},\ref{eq:neq2offdiag-ops}).

\begin{figure}[htbp]
  \includegraphics[scale=0.3,clip=true,angle=270]{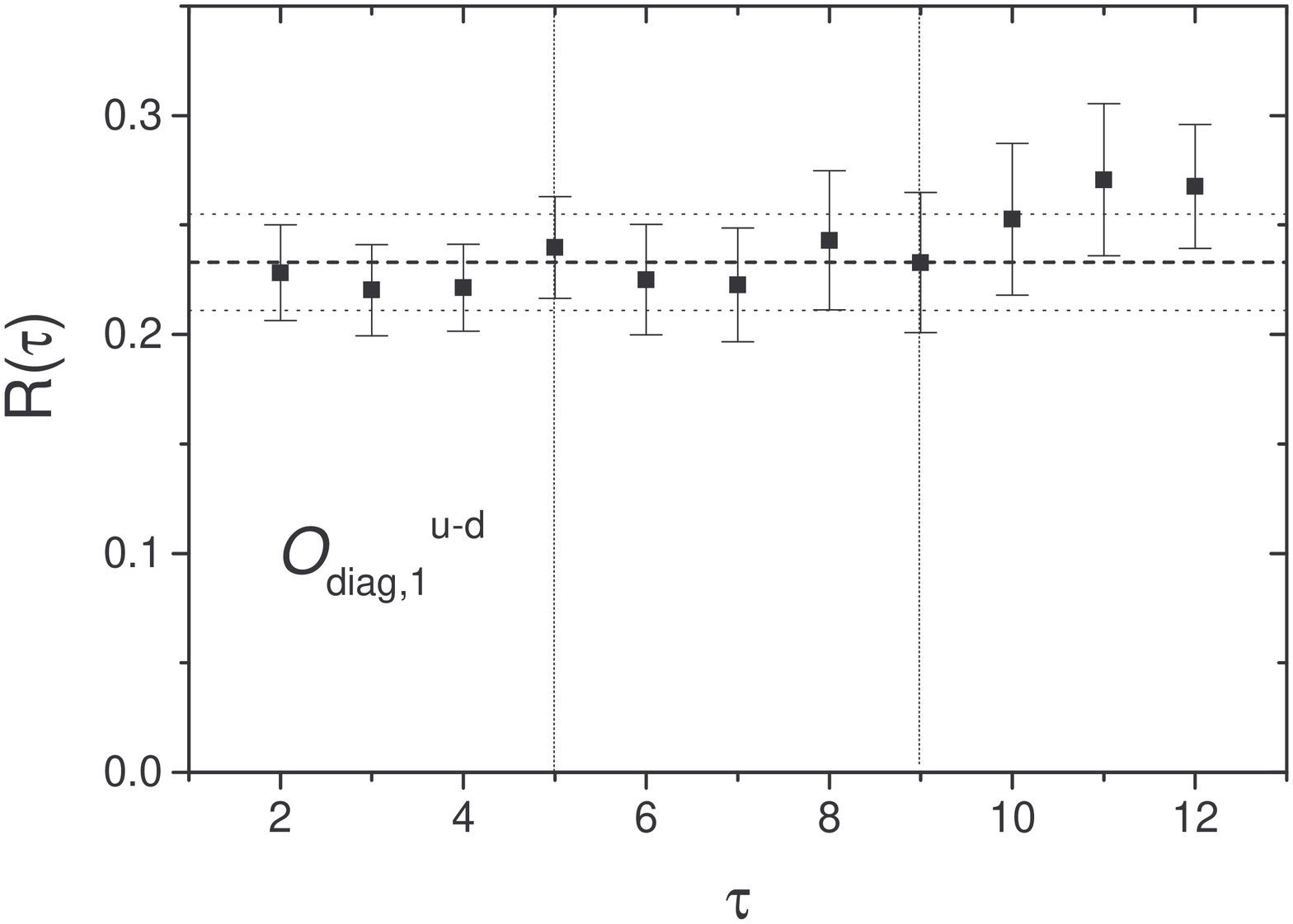}\\[-.5cm]
  \includegraphics[scale=0.3,clip=true,angle=270]{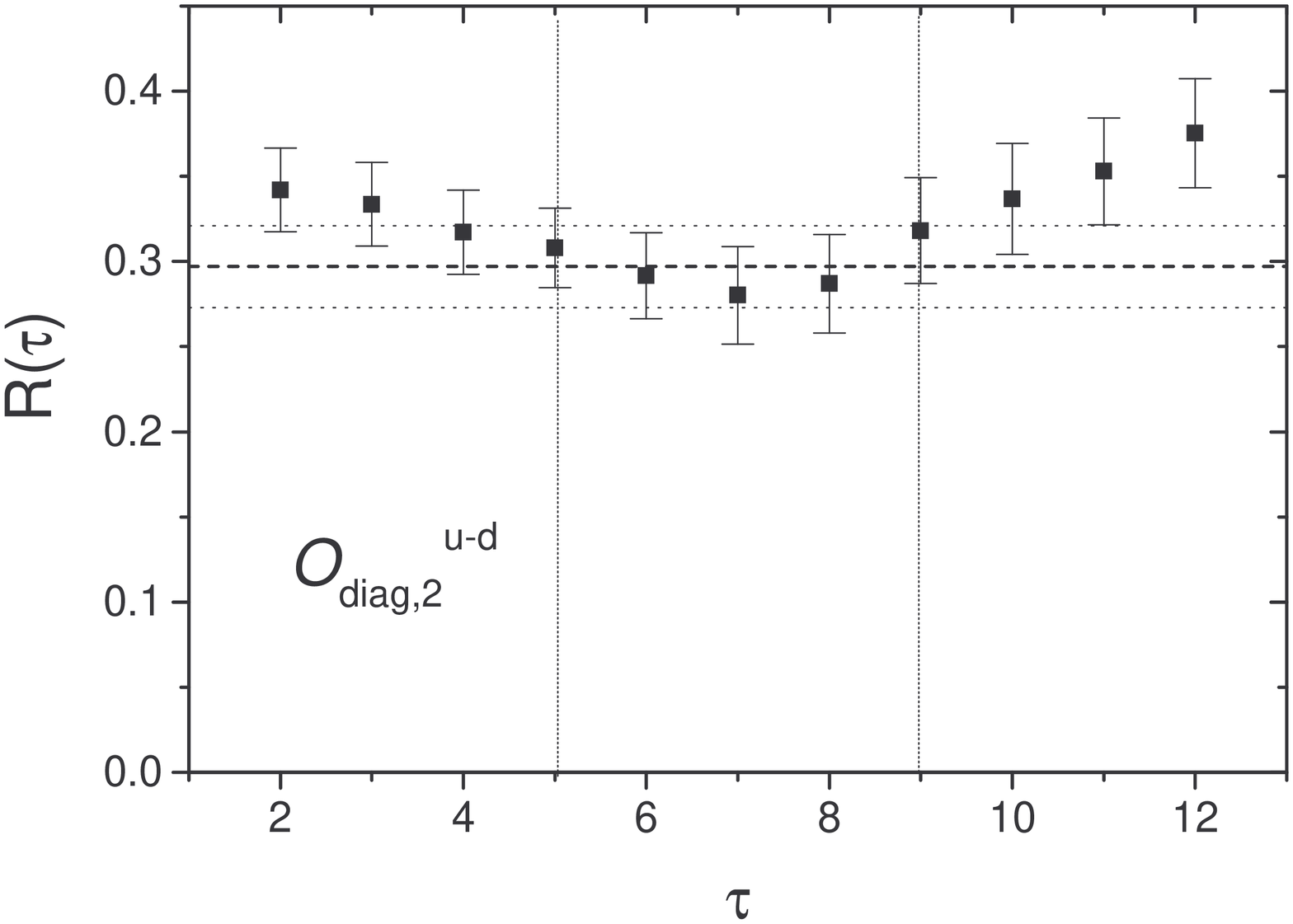}\\[-.5cm]
  \includegraphics[scale=0.3,clip=true,angle=270]{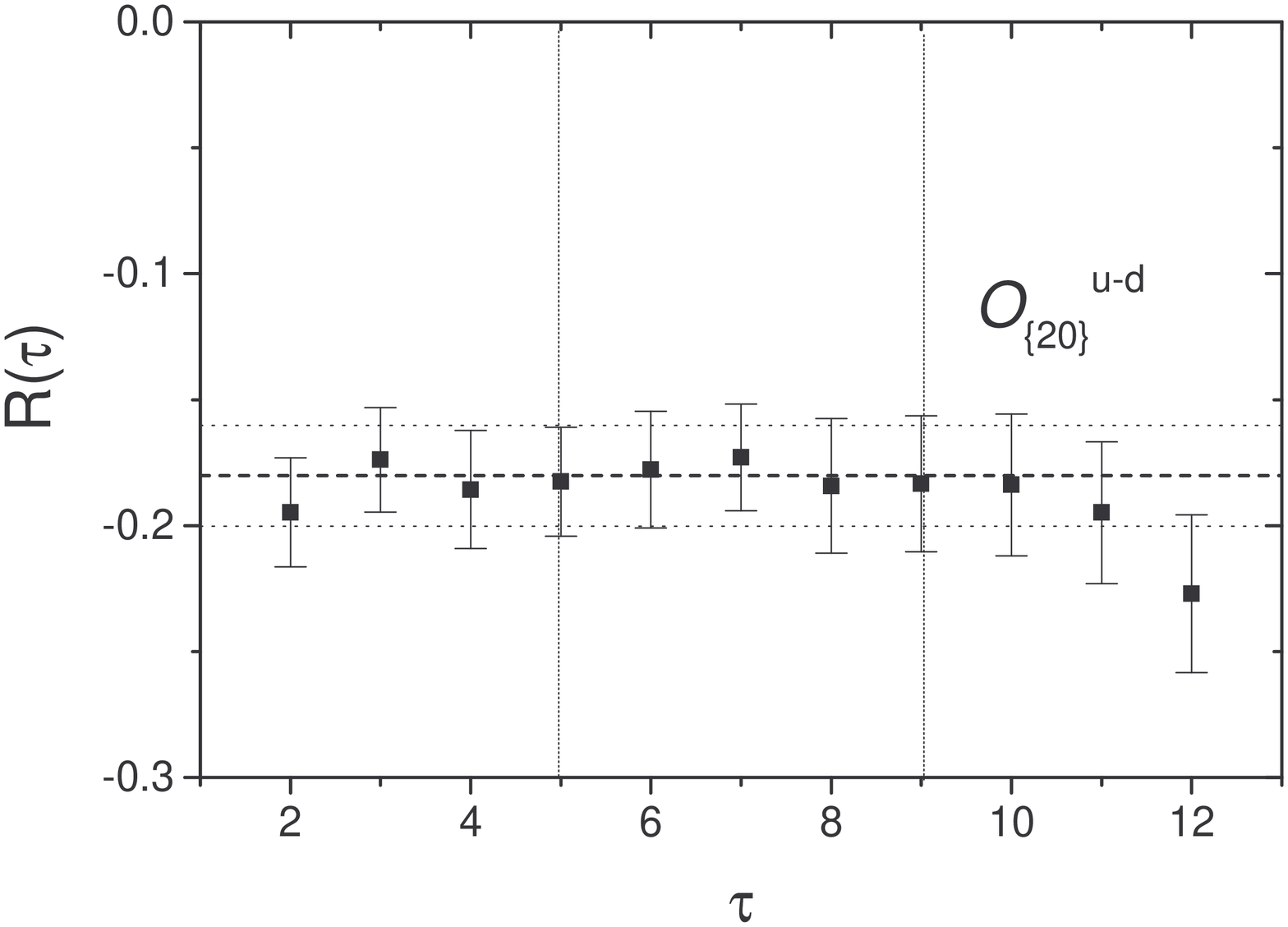}
  \caption{\label{fig:plateaus}
    Plateau plots of the ratios $R(\ts,P',P)$ for the $n = 2$
    operators ${\cal O}^{\text{u-d}}_{\text{diag,1}}$, ${\cal
      O}^{\text{u-d}}_{\text{diag,2}}$, and ${\cal
      O}^{\text{u-d}}_{\lbrace 20 \rbrace}$.}
\end{figure}

Plateau plots of the ratios, $R(\ts,P',P)$, are shown in
Fig.~\ref{fig:plateaus} for three matrix elements, $ \langle
P'\vert{\cal O}^{\text{u-d}}_{\text{diag,1}}\vert P\rangle\,, \langle
P'\vert{\cal O}^{\text{u-d}}_{\text{diag,2}}\vert P\rangle\,,$ and $
\langle P'\vert{\cal O}^{\text{u-d}}_{\lbrace 20\rbrace}\vert
P\rangle$. Measurements are averaged over the interval $\left[
  \ts_{\text{min}} = 5, \ts_{\text{max}} = 9\right]$ and this interval
is denoted by vertical lines in the plots. The solid horizontal lines
show the extracted plateau values and the dotted lines denote one
standard deviation.

The resulting lattice matrix elements, together with their jackknife
errors are as follows:
\begin{eqnarray}
  \label{eq:ratio-results}
  \overline{R}^{\text{u-d}}_{\text{diag,1}}(P^\prime,P)     &=&
  0.2327  \pm 0.0224\,, \nonumber \\
  \overline{R}^{\text{u-d}}_{\text{diag,2}}(P^\prime,P)     &=&
  0.2970  \pm 0.0235\,, \nonumber \\
  \overline{R}^{\text{u-d}}_{\text{diag,3}}(P^\prime,P)     &=&
  0.00832 \pm 0.01651\,, \nonumber \\
  \overline{R}^{\text{u-d}}_{\lbrace 10\rbrace}(P^\prime,P) &=&
  -0.09768 \pm 0.01648\,, \nonumber \\
  \overline{R}^{\text{u-d}}_{\lbrace 20\rbrace}(P^\prime,P) &=&
  -0.1802  \pm 0.0205\,, \nonumber \\
  \overline{R}^{\text{u-d}}_{\lbrace 21\rbrace}(P^\prime,P) &=&
  -0.05926 \pm 0.01162\,.
\end{eqnarray}
We now consider a minimal set of three operators to determine, but not
overdetermine, the three generalized form factors,
$A^{\text{u-d}}_{20}, B^{\text{u-d}}_{20}$, and $C^{\text{u-d}}_{20}$.
The subset of diagonal operator equations alone are insufficient
\cite{Schroers:2002ta}, since the diagonal index combinations are
linearly dependent and provide only a system of equations of rank two.

The minimal choice to determine, but not overdetermine, the form
factors is to choose two diagonal operators and one off-diagonal
operator, analogous to familiar procedures to calculate moments of
parton distributions\cite{Gockeler:1996wg,Dolgov:2002zm} or
electromagnetic form factors \cite{Gockeler:2001us,Gockeler:2003ay}.
The best choice for this purpose is the set of matrix elements
$\langle P'\vert{\cal O}^{\text{u-d}}_{\text{diag,1}} \vert P\rangle$,
$\langle P'\vert{\cal O}^{\text{u-d}}_{\text{diag,2}} \vert P\rangle$,
and $\langle P'\vert{\cal O}^{\text{u-d}}_{\lbrace 20\rbrace}\vert
P\rangle$, which have the smallest relative errors and provide a
system of three linearly independent equations.

\begin{figure}
  \includegraphics[scale=0.3,clip=true,angle=270]{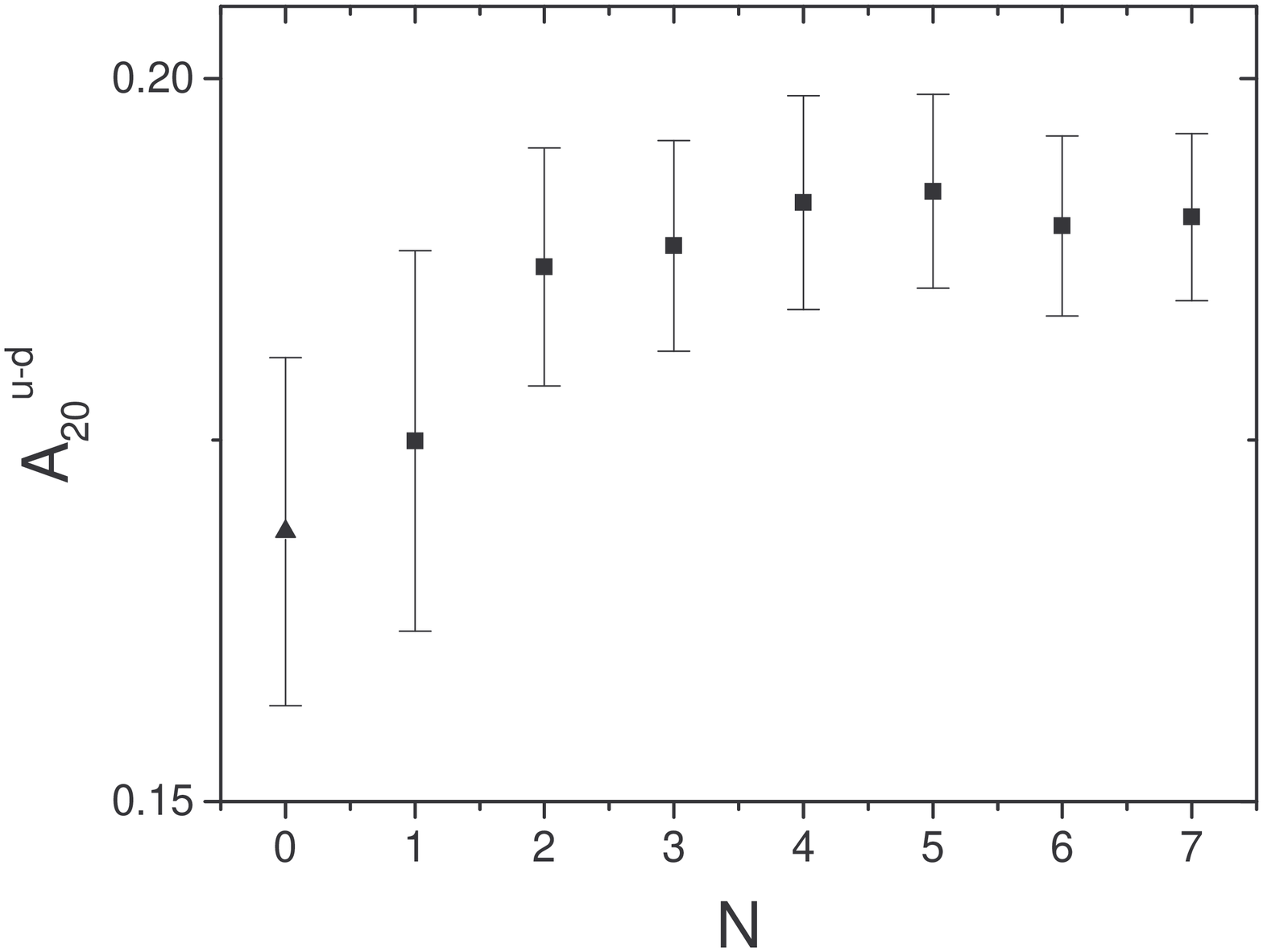} \\[-.5cm]
  \includegraphics[scale=0.3,clip=true,angle=270]{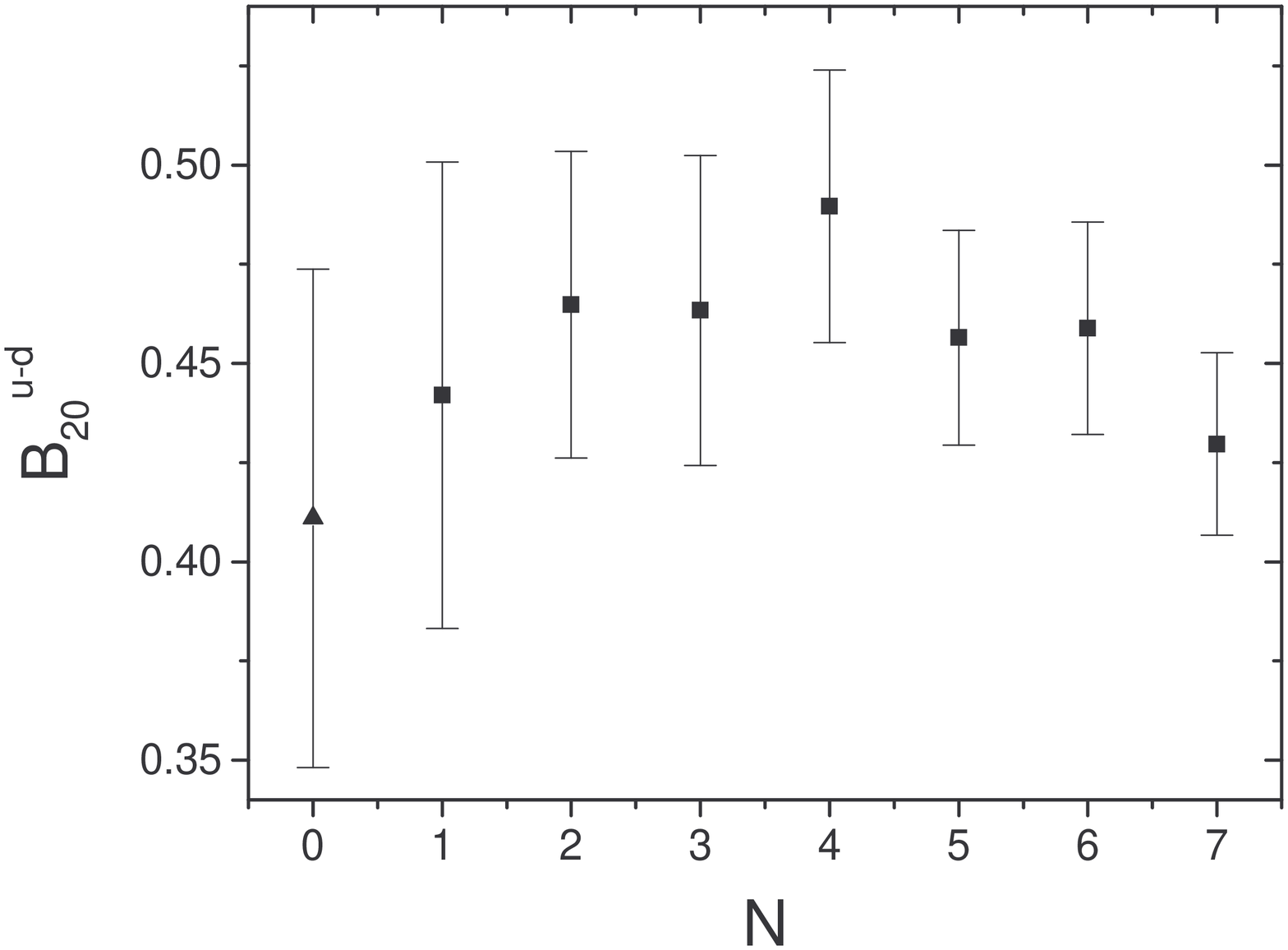}\\[-.5cm]
  \includegraphics[scale=0.3,clip=true,angle=270]{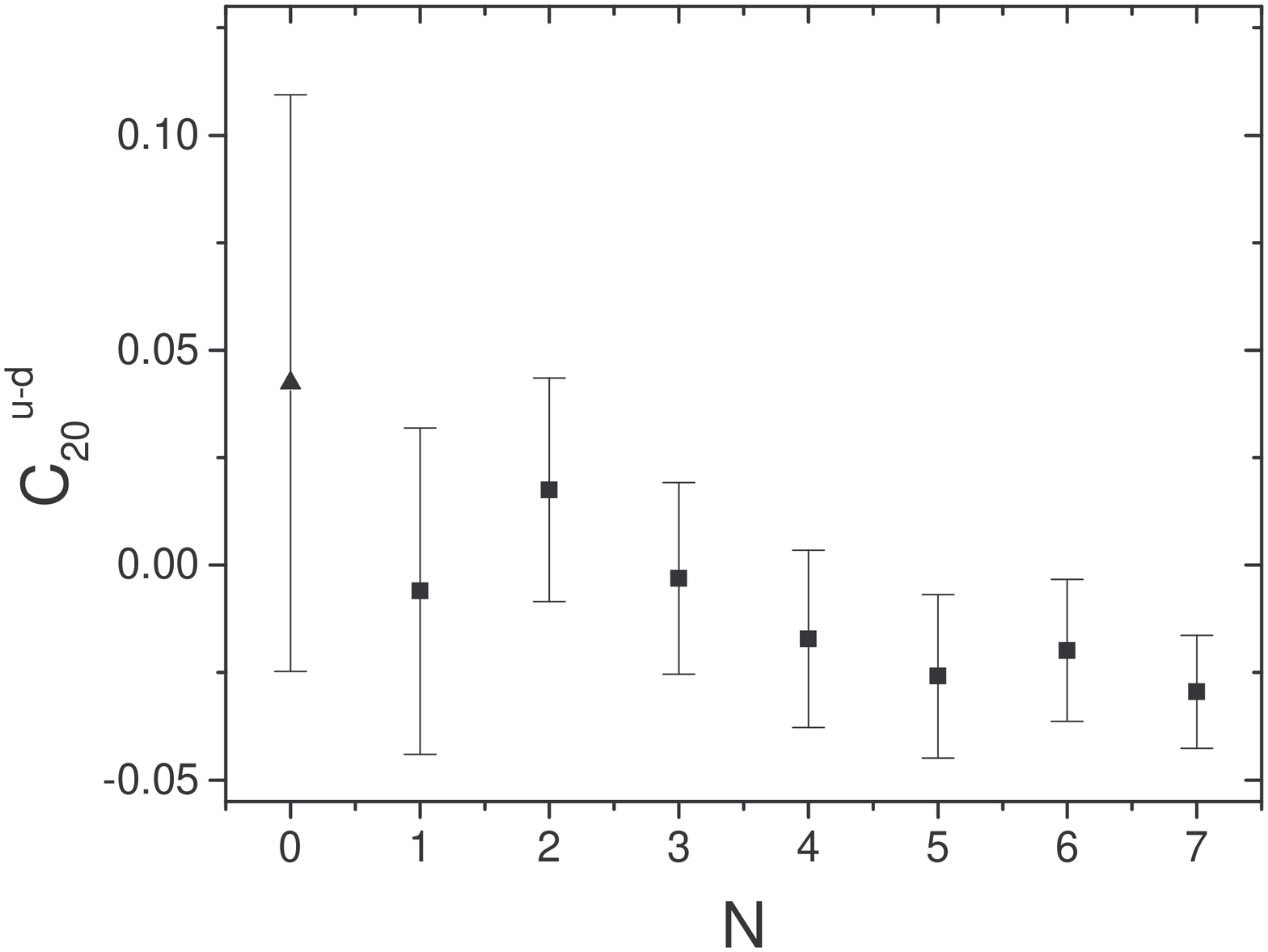}
  \caption{\label{fig:multi-mom-fit}Generalized 
    form factors obtained by simultaneous fits to $N$ external
    momentum combinations having virtuality $t_{\text{vlow}}$. As
    described in the text, the $N = 0$ points, denoted by triangles,
    use three operators at a single external momentum combination to
    determine the three form factors.  The remaining points, denoted
    by squares, use six operators and $N$ external momentum
    combinations to determine the three form factors.}
\end{figure}

The form factors and associated errors determined this way are shown
in Fig.~\ref{fig:multi-mom-fit} by the triangles plotted to at the
left at $N=0$.

Still considering the same single choice of external momenta, the full
set of six matrix elements in Eq.~(\ref{eq:gff-oplist}) provides an
overdetermined set of equations that yield the form factors and errors
labeled $N=1$ in Fig~\ref{fig:multi-mom-fit}. We previously argued
that the $C$ form factor is least accurately determined because its
coefficients have the highest powers of $\Delta$, and the minimally
determined fit, $N=0$ bears out this expectation. Note that the
overdetermined fit, $N=1$ leaves the errors in $A$ and $B$ roughly the
same, but substantially reduces that of $C$. The results of the two
fits agree within errors, but since the overdetermined fit exploits
the maximal information in the lattice measurements,
Eq.~(\ref{eq:ratio-results}) and has smaller overall statistical
errors, we consider it superior.

We now progress to the next level by including all the external
momentum combinations enumerated in Tab.~\ref{tab:low-virt}. The
results are displayed in Fig.~\ref{fig:multi-mom-fit}, where the
abscissa denotes the total number of momentum combinations included in
the fit. Entry $N$ corresponds to the inclusion of momentum sets $1, 2,
\cdots N$ of Tab.~\ref{tab:low-virt}. As expected, the errors for each
form factor decrease significantly as new statistical information is
included by adding momenta in new directions. The improvement is
generally less than $\sqrt N$ because measurements of different
momenta on the same lattices are correlated.

The overall result in Fig.~\ref{fig:multi-mom-fit} is a strong
validation of our approach. Comparing the minimally determined, $N=0$
result at the far left with the overdetermined result with all seven
momentum combinations, $N=7$ at the far right, we observe that our
method reduces the error in the least accurately determined form
factor $C$ by a factor of 5, the error in $B$ by a factor of 3, and
the error in $A$ by a factor of 2.

\subsection{\label{sec:quark-ang-mom}Generalized Form Factors and 
  Quark Angular Momentum} 
\begin{figure}[htbp]
  \begin{center}
    \includegraphics[scale=0.5,clip=true,angle=270]{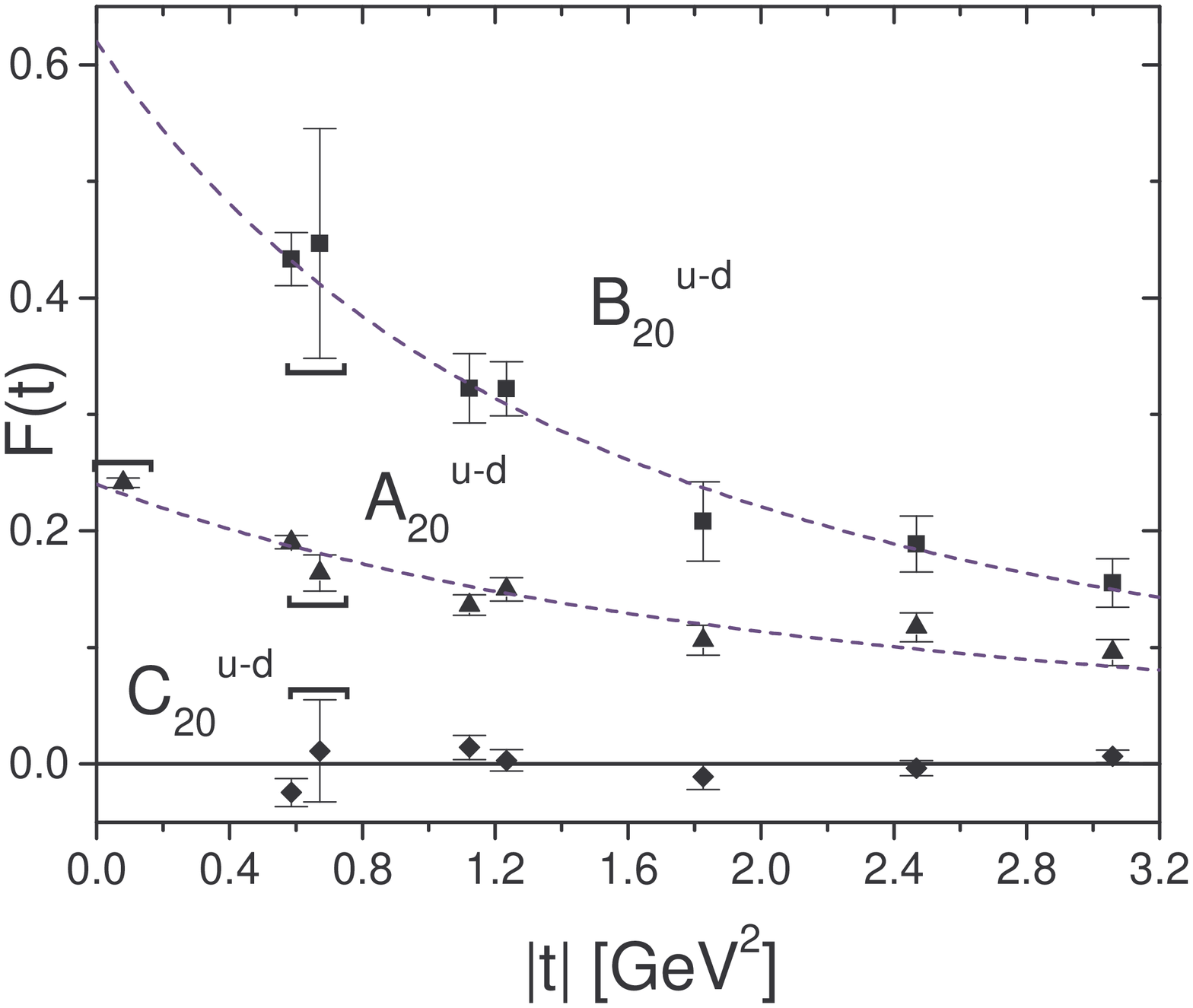}
    \caption{\label{fig:n2-fu-d}Generalized form factors
      $A^{\text{u-d}}_{2 0}(t)$, $B^{\text{u-d}}_{2 0}(t)$, and
      $C^{\text{u-d}}_{2}(t)$ for all available virtualities
      obtained using the full set of operators and external momentum
      combinations.  The dashed curves denote dipole fits to $A$ and
      $B$ to guide the eye and extrapolate to $t = 0$. The form factor
      $C$ is consistent with zero.  Four data points, denoted by
      horizontal brackets, have been shifted by $\vert t\vert=0.2\,$GeV$^2$
      to the right for
      clarity in plotting.}
  \end{center}
\end{figure}

Having validated our analysis technique, we
now apply it to the full set of external momenta and virtualities
listed in Appendix~\ref{sec:matr-elem-terms},
Tab.~\ref{tab:all-latt-moms}. The results of the full, overdetermined
analysis for the flavor non-singlet generalized form factors
$A^{\text{u-d}}_{2 0}(t)$, $B^{\text{u-d}}_{2 0}(t)$, and
$C^{\text{u-d}}_{2}(t)$ are shown in Fig.~\ref{fig:n2-fu-d}. We
observe that with the present analysis, all three generalized form
factors are determined quite accurately throughout the full range of
virtuality up to $t = 3.1\,$GeV$^2$. As expected, some combinations of
external momenta induce more statistical noise than others, but the
overall structure of the form factors is well determined. Although
there is no fundamental argument for the functional dependence on $t$,
it is useful to fit the form factors to the dipole form that provides
a good phenomenological fit to the nucleon electromagnetic form
factors, and the results of least-squares fit by dipole form factors
are shown for $A(t)$ and $B(t)$. Although the individual up and down
form factors $C_2^{\text{u/d}}(t)$ are non-vanishing, their
difference, $C_2^{\text{u-d}}(t)$, is statistically consistent with
zero. We note this flavor independence is a feature of the Chiral
Quark Soliton Model in Ref.~\cite{Kivel:2000fg}.

Although results are not yet available at other quark masses to enable
extrapolation to the physical quark mass, these results clearly show
the behavior of the generalized form factors in the $900\,$MeV pion
world.  For this non-singlet case, there are no corrections from
disconnected diagrams.

It is useful to point out that in all cases for which $t \neq 0$,
there were no zero (or nearly zero) singular values in our
calculations.  Hence, in these cases, the singular value decomposition
is completely equivalent to minimizing Eq.~(\ref{eq:chisq-min}) by
conventional least-squares analysis.  In the case of $t = 0$, because
of the explicit factors of $\Delta$ appearing in
Eq.~(\ref{eq:explicit}), only the coefficients $A^q_{n0}(0)$ can be
determined.  As expected, the singular value decomposition handled
this automatically, with the zero singular values resulting in
minimization in the appropriate subspace of $A^q_{n0}(0)$. Note that
this procedure introduces no bias or model assumptions. Rather, it
simply specifies the correct physical subspace.

\begin{figure}[htbp]
  \begin{center}
    \includegraphics[scale=0.5,clip=true,angle=270]{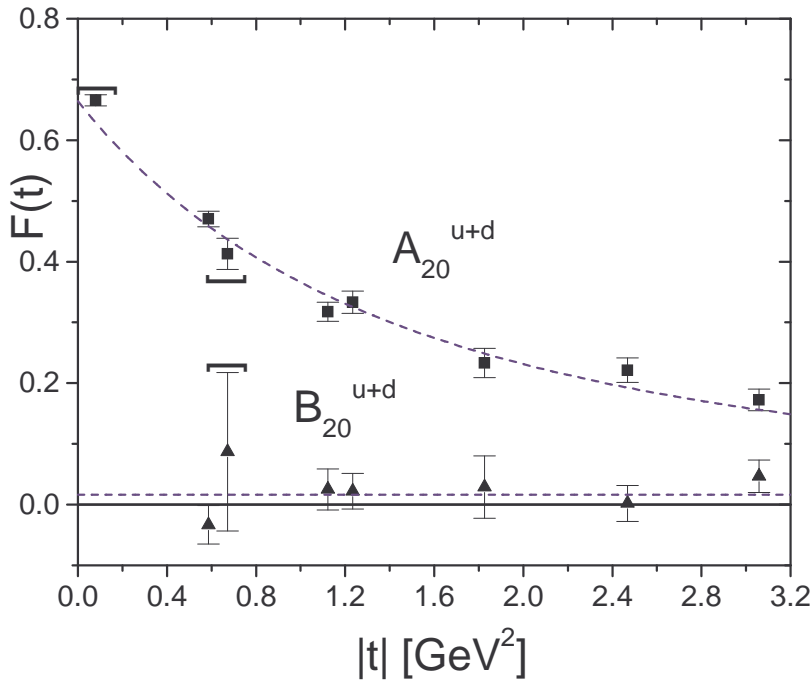}
    \caption{Flavor singlet generalized form factors
      $A^{\text{u+d}}_{2 0}(t)$ and $B^{\text{u+d}}_{2 0}(t)$ with
      dipole fits denoted by dashed curves. Note that the singlet
      combination $B^{\text{u+d}}_{2 0}(t)$ is consistent with zero,
      so that the total quark angular momentum $J_q = \frac{1}{2}
      [A^{\text{u+d}}_{2 0}(0) + B^{\text{u+d}}_{2 0}(0) ]$ is
      dominated by $A$. Three data points, denoted by horizontal
      brackets, have been shifted by $\vert t\vert=0.2\,$GeV$^2$ to
      the right for clarity in plotting.}
    \label{fig:n2-fupd}
  \end{center}
\end{figure}

Since the total quark angular momentum is given by the zero virtuality
limit, $J_q = \frac{1}{2} [A^{\text{u+d}}_{2 0}(0) + B^{\text{u+d}}_{2
  0}(0) ]$, it is particularly interesting to study this limit. Two
issues need to be addressed. The first is extrapolation to $t = 0$.
This is not a problem for $A$, since it is just the moment of the
spin-averaged parton distribution $A_{2 0}(0) = \langle x \rangle $.
However, like the electromagnetic form factor $F_2$, $B$ cannot be
measured at $t=0$ since the kinematical factors always contain
$\Delta$, and therefore it must be extrapolated. From the behavior of
the flavor non-singlet combination $B^{\text{u-d}}_{2 0}(t)$ in
Fig.~\ref{fig:n2-fu-d}, one might expect significant extrapolation
errors. However, as shown in Fig.~\ref{fig:n2-fupd}, the cancellation
between the up and down quark contributions is nearly complete and the
flavor singlet combination $B^{\text{u+d}}_{2 0}(t)$ is consistent
with zero. Hence, essentially the full contribution comes from
$A^{\text{u+d}}_{2 0}(0)$.

The second issue is the fact that disconnected diagrams, which are
beyond the scope of the present work, must also be included in these
flavor singlet matrix elements.

It is interesting to note that light cone arguments indicate that the
contribution of each Fock space sector to the gravitomagnetic moment
$B(0) = \sum_{q,g} B_{20}^{q,g}(0) = 0$ is separately
zero~\cite{Brodsky:2000ii}. Although this may not necessarily imply
that connected and disconnected diagrams for $B(0)$ separately cancel,
it is suggestive of our result above that the connected contribution
to $B(0)$ was consistent with zero. Note that the separate vanishing
of quark and gluon contributions to $B(0)$ was also conjectured in
Ref.~\cite{Teryaev:1998iw}.

With the caveat that we must omit disconnected diagrams for the
present, we proceed to quote our lattice results from connected
diagrams. Using the results of Ref.~\cite{Dolgov:2002zm} at $\kappa =
0.1560$, the quark spin contribution to nucleon angular momentum is
given by the lowest moment of the longitudinal spin distribution
\begin{eqnarray}
  \frac{1}{2} \Delta \Sigma &= & \frac{1}{2} [\langle 1 \rangle_{\Delta 
    u} +\langle 1 \rangle_{\Delta d} ]  \nonumber \\
  &\sim &\frac{1}{2} 
  \, \, 0.682(18)\,.
\end{eqnarray}
Since $B^{\text{u+d}}_{2 0}(0) \sim 0$, the total quark spin and
orbital contribution to the nucleon angular momentum is
\begin{eqnarray}
  J_q & = & \frac{1}{2} [A^{\text{u+d}}_{2 0}(0) + B^{\text{u+d}}_{2 
    0}(0) ]  \nonumber \\ 
  & \sim & \frac{1}{2}[\langle x \rangle_{ u} 
  +\langle x \rangle_{ d} + 0 ]
  \nonumber \\ 
  & \sim & \frac{1}{2} \, 
  \, 0.675(7)\,.
\end{eqnarray}
Thus, since these two results are equal within errors, if we consider
only connected diagrams in the world of $900\,$MeV pions, we conclude
that the quark spin produces 68 percent of the nucleon angular
momentum and the quark orbital angular momentum contribution is
negligible.  Although this is superficially consistent with
expectations from relativistic quark models, we note that serious
physical interpretation and quantitative comparison with quark models
requires consideration of the disconnected diagrams.  The striking
fact for our present purposes, however, is that it is possible to
measure the connected part to the order of one percent, under the
assumption that $B^{\text{u+d}}_{20}(0)=0\pm 0$.

Note that quenched connected diagram contributions of $\Delta \Sigma$
and $J_q$ have been reported in
Refs.~\cite{Mathur:1999uf,Gockeler:2003ar}. Since those calculations
were performed for pion masses ranging above $500\,$MeV, we would
expect quenching effects to be negligible and hence that their results
would be statistically consistent with ours. The chiral extrapolated
results of G{\"o}ckeler et.~al.~\cite{Gockeler:2003ar} that 66 $\pm$ 14
percent of the nucleon spin arises from the quark spin and that 6 $\pm$
14 percent arises from quark orbital angular momentum are completely
compatible with our results because of the small quark mass
dependence.  The results for $\Delta \Sigma$ by Mathur et.~al.~in
Ref.~\cite{Mathur:1999uf} are also consistent with ours.
Reference~\cite{Dolgov:2002zm} showed that the linear extrapolation of
$\Delta \Sigma$ is essentially constant, only changing from 68 $\pm$ 2
percent of the nucleon spin at $m_{\pi} = 900\,$ MeV to 69 percent in
the chiral limit.  This is consistent with the connected diagram
chiral extrapolation by Mathur et.~al.~of 62 $\pm$ 8 percent.  The
results for $J_q$, however, are inconsistent. From the results of
their Fig.~3 and their 1.045 renormalization factor, $J_q$ contributes
84 $\pm$ 8 percent of the spin, which disagrees with our result of 68
$\pm$ 1 percent.  We believe the discrepancy arises because Mathur
et.~al.~use the dipole prescription to extrapolate the sum
$A_{20}^{\text{u+d}} + B_{20}^{\text{u+d}}$, whereas, as shown in our
Fig.~\ref{fig:n2-fupd}, we calculate $A^{\text{u+d}}_{20}(0)$ directly
without extrapolation and only extrapolate the nearly vanishing
$B^{\text{u+d}}_{20}$.

%
%

\section{\label{sec:Conclusions}Conclusions}
In summary, we have presented a new method for calculating generalized
form factors in lattice QCD that extracts the full information content
from a given lattice configuration by measuring an overdetermined set
of lattice observables. We demonstrated its effectiveness in an
exploratory calculation of $n=2$ generalized form factors up to
$3\,$GeV$^2$ and showed that it reduces errors to as small as
one-fifth of those obtained by a conventional minimally-determined
analysis. The final error bars for $A^{\text{u-d}}_{2 0}(t)$,
$A^{\text{u+d}}_{2 0}(t)$, and $B^{\text{u-d}}_{2 0}(t)$ are typically
of the order of five to ten percent, providing useful information
about the Fourier transform of the transverse structure of the
nucleon. Because of fortunate cancellation in the flavor singlet case,
the connected diagram contributions to the total quark angular
momentum are measured to the order of one percent.  Although this
exploratory calculation was performed for the heaviest of the three
SESAM quark masses used in earlier calculation of moments of parton
distributions, we expect the technique to be sufficiently robust to
treat all the masses.

We note that the first calculation of generalized form factors were
reported by Schroers in~\cite{Schroers:2002ta} using the standard
minimally-determined analysis, and more extensive calculations by the
QCDSF collaboration using our method are being published
simultaneously with this present work~\cite{Gockeler:2003ar}.

This work provides the foundation for a number of promising
investigations. The $n=3$ relations introduced in this work enable us
to calculate the $t$-dependence for three moments and thus explore the
variation of the transverse structure of the proton with $x$. The same
methodology can, of course, be used to calculate spin-dependent
generalized form factors. In the longer term, extensive calculations
with emerging multi-Teraflops computers dedicated to lattice QCD will
enable the chiral and continuum calculations required to have
definitive impact on the new generation of experiments currently being
undertaken~\cite{Airapetian:2001yk,Stepanyan:2001sm,Favart:2002eh}.

%
%

\begin{acknowledgments}
  The authors wish to thank Matthias Burkardt, Meinulf G{\"o}ckeler,
  Axel Kirchner, Andreas Sch{\"a}fer, Oleg V.~Teryaev and Christian
  Weiss for stimulating discussions. P.H.~and W.S.~are grateful for
  Feodor-Lynen Fellowships from the Alexander von Humboldt Foundation
  and thank the Center for Theoretical Physics at MIT for its
  hospitality. This work is supported in part by the U.S.~Department
  of Energy (D.O.E.) under cooperative research agreement
  \#DF-FC02-94ER40818.
\end{acknowledgments}

%
%

\appendix

\section{\label{sec:matr-elem-terms}Matrix elements in terms of
  generalized form factors}
In order to find the relation between Minkowski-expressions, e.g.
$\left( \mathcal{O}^{\text{M}}\right) ^{\{\mu _{1}\mu
  _{2}\dots\mu_{n}\}}$, and the corresponding Euclidean quantities,
$\left( \mathcal{O}^{\text{E}}
\right)_{\{\mu_{1}\mu_{2}\dots\mu_{n}\}}$, we observe that the
operators under consideration are constructed from gamma matrices and
covariant derivatives $D$.
The conventions we use, consistent with Ref.~\cite{Montvay:1994cy},
are
\begin{eqnarray*}
  \left( \gamma ^{\text{E}}\right) _{\mu } &=&\sum_{\nu =1}^{4}h_{\mu
    \nu
  }\left( \gamma ^{\text{M}}\right) ^{\nu },\quad h_{\mu \nu
  }=\text{diag}%
  \left( \text{i},\text{i},\text{i},1\right) , \\
  \left( D^{\text{E}}\right) _{\mu } &=&\sum_{\nu =1}^{4}d_{\mu \nu
  }\left( D^{%
      \text{M}}\right) ^{\nu },\quad d_{\mu \nu }=\text{diag}\left(
    1,1,1,-\text{i}%
  \right)\,,
\end{eqnarray*}
and we will set $\left( a^{\text{M}}\right) ^{\nu =0}=$ $\left(
  a^{\text{M}}\right) ^{\nu =4}$. Since all operators are symmetrized,
the general transformation rule for the operators with one gamma
matrix and $n-1$ covariant derivatives investigated in this work is
given by
\begin{equation}
  \left( \mathcal{O}^{\text{E}}\right) _{\{\mu _{1}\mu _{2}\dots\mu
    _{n}\}}=h_{\mu _{1}\nu _{1}}d_{\mu _{2}\nu _{2}}\dots d_{\mu _{n}\nu
    _{n}}\left( \mathcal{O}^{\text{M}}\right) ^{\{\nu _{1}\nu _{2}\dots\nu
    _{n}\}},
  \label{eq:MEtransf}
\end{equation}
where a summation over the indices $\nu _{i}$ is implicit. In the
following, we will drop the label E or M, and it will be clear from
the context if Euclidean or Minkowski expressions are used.

We write matrix elements in terms of Dirac spinors as follows
\[
\left\langle P^{\prime }\right| \mathcal{O}\left| P\right\rangle
=\overline{U}(P^{\prime }){\mathcal{K}}_{\mathcal{O}}(P^{\prime
},P)U(P)\,.
\]
The ratio $R_{\mathcal{O}}^{\text{cont}}(P^{\prime },P)$ can be
written explicitly in Minkowski space as
\begin{eqnarray}
  \left. R_{\mathcal{O}}^{\text{cont}}(P^{\prime },P)\right|
  _{\text{ground
      state}} &=&\left( E(P^\prime)E(P)\right) ^{-1/2} \nonumber \\
  && \quad\times \left(
    \frac{1}{2}\text{Tr}%
    \left[ \Gamma _{\text{unpol}}\left( \not{P}^{\prime }+m\right)
    \right] \frac{%
      1}{2}\text{Tr}\left[ \Gamma _{\text{unpol}}\left( \not{P}+m\right)
    \right]
  \right) ^{-1/2}  \nonumber \\
  &&\quad\times\frac{1}{4}\text{Tr}\left[ \Gamma _{\text{pol}}\left(
      \not{P}^{\prime
      }+m\right) \mathcal{K}_{\mathcal{O}}(P^{\prime },P)\left(
      \not{P}+m\right)
  \right]\,,
  \label{eq:RatioM}
\end{eqnarray}
neglecting all excited states which would introduce a time dependence.
Note that $R_{\mathcal{O}}^{\text{cont}}(P^{\prime },P)$ is directly
proportional to $\left\langle P^{\prime }\right| \mathcal{O}\left|
  P\right\rangle $. The ratio in Eq.~(\ref{eq:RatioM}) is expressed in
a general form that applies to all quark bilinear operators and
projectors $\Gamma _{\text{pol/unpol}}$ relevant to this work.

The Minkowski space expressions for the QCDSF and the LHPC projectors
used in our analysis are as follows
\begin{eqnarray}
  \Gamma _{\text{unpol}}^{\text{LHPC}} &=&\frac{1}{4}\left( 1+\gamma
    _{0}\right) ,\;\Gamma _{\text{pol}}^{\text{LHPC}}=\frac{1}{4}\left(
    1+\gamma
    _{0}\right) \left( 1-\gamma _{5}\gamma _{3}\right)\,,
  \label{eq:LHPC_proj} \\
  \Gamma _{\text{unpol}}^{\text{QCDSF}} &=&\frac{1}{2}\left( 1+\gamma
    _{0}\right) ,\;\Gamma _{\text{pol}}^{\text{QCDSF}}=\frac{1}{2}\left(
    1+\gamma _{0}\right) \gamma _{5}\gamma _{2}\,.
  \label{eq:QCDSF_proj}
\end{eqnarray}
We also record the complete expressions for the kernels $%
\mathcal{K}_{\mathcal{O}}(P^{\prime },P)$ in Eq.~(\ref{eq:RatioM}).
Following the general parameterization of \cite{Ji:1997ek}, we have for
the lowest three moments $n=1,2,3$ in Minkowski space
\begin{eqnarray}
  \left\langle P^{\prime }\right| \mathcal{O}^{\{\mu _{1}\mu
    _{2}\dots\}}\left| P\right\rangle  &=&\overline{U}(P^{\prime
  })\mathcal{K}_{%
    \mathcal{O}}^{\{\mu _{1}\mu _{2}\dots\}}(P^{\prime },P)U(P)
  \nonumber
  \\
  \left\langle P^{\prime }\right| \overline{\psi }_{q}\gamma ^{\mu
    _{1}}\psi _{q}\left| P\right\rangle  &=&\overline{U}(P^{\prime
  })\left\{
    \gamma ^{\mu _{1}}A_{q,1}(\Delta ^{2})+\text{i}\frac{\sigma ^{\mu
        _{1}\alpha
      }\Delta _{\alpha }}{2m}B_{q,1}(\Delta ^{2})\right\} U(P)
  \nonumber \\ 
  \left\langle P^{\prime }\right| \overline{\psi }_{q}\gamma ^{\{\mu
    _{1}}%
  \overleftrightarrow{D}^{\mu _{2}\}}\psi _{q}\left| P\right\rangle
  &=&%
  \overline{U}(P^{\prime })\left\{ (-\text{i})\overline{P}^{\{\mu
      _{1}}\gamma
    ^{\mu _{2}\}}A_{q,2}(\Delta ^{2})\right.   \nonumber \\
  &&+(-\text{i})\text{i}\overline{P}^{\{\mu _{1}}\frac{\sigma ^{\mu
      _{2}\}\alpha }\Delta _{\alpha }}{2m}B_{q,2}(\Delta ^{2})
      \nonumber \\ 
  &&+\left. (-\text{i})\frac{\Delta ^{\{\mu _{1}}\Delta ^{\mu
        _{2}\}}}{m}%
    C_{q,2}(\Delta ^{2})\right\} U(P)  \nonumber \\
  \left\langle P^{\prime }\right| \overline{\psi }_{q}\gamma ^{\{\mu
    _{1}}%
  \overleftrightarrow{D}^{\mu _{2}}\overleftrightarrow{D}^{\mu
    _{3}\}}\psi
  _{q}\left| P\right\rangle  &=&\overline{U}(P^{\prime })\left\{
    (-\text{i}%
    )^{2}\overline{P}^{\{\mu _{1}}\overline{P}^{\mu _{2}}\gamma ^{\mu
      _{3}\}}A_{q,30}(\Delta ^{2})\right.   \nonumber \\
  &&+(-\text{i})^{2}\Delta ^{\{\mu _{1}}\Delta ^{\mu _{2}}\gamma ^{\mu
    _{3}\}}A_{q,32}(\Delta ^{2})  \nonumber \\
  &&+\text{i}(-\text{i})^{2}\overline{P}^{\{\mu _{1}}\overline{P}^{\mu
    _{2}}%
  \frac{\sigma ^{\mu _{3}\}\alpha }\Delta _{\alpha
    }}{2m}B_{q,30}(\Delta ^{2})
  \nonumber \\
  &&\left. +\text{i}(-\text{i})^{2}\Delta ^{\{\mu _{1}}\Delta ^{\mu
      _{2}}\frac{%
      \sigma ^{\mu _{3}\}\alpha }\Delta _{\alpha }}{2m}B_{q,32}(\Delta
    ^{2})\right\} U(P)\,.
  \label{eq:parametrization}
\end{eqnarray}
In practice, we evaluate the ratio $R_{\mathcal{O}}^{\text{cont,}\{\mu
  _{1}\mu _{2}\dots\}}(P^{\prime },P)$ by inserting the kernels
$\mathcal{K}_{\mathcal{O}}^{\{\mu _{1}\mu _{2}\dots\}}$ from
Eq.~(\ref{eq:parametrization}) and the projectors in
Eq.~(\ref{eq:LHPC_proj},\ref{eq:QCDSF_proj}) into
Eq.~(\ref{eq:RatioM}). For given values of the nucleon momenta $P$ and
$P^{\prime }$, the ratio is computed numerically, determining the
kinematical coefficients of the GFFs $A(t)$, $B(t)$ and $C(t)$.
Eventually, we transform the free indices $\{\mu _{1}\mu _{2}\dots\}$
to Euclidean space with the use of Eq.~(\ref{eq:MEtransf}). The system
of linear equations is then constructed by equating the Euclidean
lattice result and the transformed Minkowskian parameterization.

\subsection{\label{sec:lattice-operators}Lattice operators}
On the lattice side, we have H$ \left( 4\right) $-invariance instead
of Lorentz-invariance, which complicates the classification and the
renormalization of the operators.  Following~\cite{Gockeler:1996mu},
we choose appropriate linear combinations of the operators
$\mathcal{O}_{\{\mu _{1}\mu _{2}\dots\}}$, corresponding to
representations of $H(4)$.

For the case of one derivative, $n=2$, there are three diagonal
operators
\begin{eqnarray}
  \label{eq:neq2diag-ops}
  \mathcal{O}_{1}^{\text{diag,n=2}} &=&\frac{1}{2}\left[
    \mathcal{O}_{11}+\mathcal{O}_{22}-\mathcal{O}_{33}-\mathcal{O}_{44}\right]
    \,,\nonumber \\
  \mathcal{O}_{2}^{\text{diag,n=2}} &=&\frac{1}{2^{1/2}}\left[
    \mathcal{O}_{33}-\mathcal{O}_{44}\right] , \nonumber \\
  \mathcal{O}_{3}^{\text{diag,n=2}} &=&\frac{1}{2^{1/2}}\left[
    \mathcal{O}_{11}-\mathcal{O}_{22}\right]\,,
\end{eqnarray}
which are symmetric and traceless by construction. Additionally there
are six non-diagonal combinations
\begin{equation}
  \label{eq:neq2offdiag-ops}
  \mathcal{O}_{\mu _{1},\mu
      _{2}}^{\text{non-diag,n=2}}=\frac{1}{2}\left[ \mathcal{O}_{\mu 
      _{1}\mu
      _{2}}+\mathcal{O}_{\mu _{2}\mu _{1}}\right]
      =\mathcal{O}_{\left\{ \mu _{1}\mu _{2}\right\}
  },\quad \mu _{1,2}=1\dots 4;\;\mu _{1}<\mu _{2}\,,
\end{equation}
giving in the most general case a set of nine independent operators.
There is no operator mixing for $n=2$.

In the case of two derivatives, $n=3$, we take
\begin{eqnarray}
  \mathcal{O}_{1}^{\text{n=3}} &=&\left( \frac{3}{2}\right) ^{1/2}\left[
    \mathcal{O}_{\left\{
        122\right\} }-\mathcal{O}_{\left\{ 133\right\} }\right] ,\quad
  \mathcal{O}_{2}^{\text{n=3}}=%
  \frac{1}{2^{1/2}}\left[ \mathcal{O}_{\left\{ 122\right\}
        }+\mathcal{O}_{\left\{
        133\right\}
    }-2\mathcal{O}_{\left\{ 144\right\} }\right] ,  \nonumber \\
  \mathcal{O}_{3}^{\text{n=3}} &=&\left( \frac{3}{2}\right) ^{1/2}\left[
    \mathcal{O}_{\left\{
        211\right\} }-\mathcal{O}_{\left\{ 233\right\} }\right] ,\quad
  \mathcal{O}_{4}^{\text{n=3}}=%
  \frac{1}{2^{1/2}}\left[ \mathcal{O}_{\left\{ 211\right\}
        }+\mathcal{O}_{\left\{
        233\right\}
    }-2\mathcal{O}_{\left\{ 244\right\} }\right] ,  \nonumber \\
  \mathcal{O}_{5}^{\text{n=3}} &=&\left( \frac{3}{2}\right) ^{1/2}\left[
    \mathcal{O}_{\left\{
        311\right\} }-\mathcal{O}_{\left\{ 322\right\} }\right] ,\quad
  \mathcal{O}_{6}^{\text{n=3}}=%
  \frac{1}{2^{1/2}}\left[ \mathcal{O}_{\left\{ 311\right\} }+O_{\left\{
        322\right\}
    }-2\mathcal{O}_{\left\{ 344\right\} }\right] ,  \nonumber \\
  \mathcal{O}_{7}^{\text{n=3}} &=&\left( \frac{3}{2}\right) ^{1/2}\left[
    \mathcal{O}_{\left\{
        411\right\} }-\mathcal{O}_{\left\{ 422\right\} }\right] ,\quad
  \mathcal{O}_{8}^{\text{n=3}}=%
  \frac{1}{2^{1/2}}\left[ \mathcal{O}_{\left\{ 411\right\}
        }+\mathcal{O}_{\left\{
        422\right\}
    }-2\mathcal{O}_{\left\{ 433\right\} }\right]\, ,
  \label{eq:Opn3}
\end{eqnarray}
while the four non-diagonal combinations are denoted by
\[
\mathcal{O}_{\mu _{1},\mu _{2},\mu
    _{3}}^{\text{non-diag,n=3}}=\mathcal{O}_{\left\{ \mu 
    _{1}\mu
    _{2}\mu _{3}\right\} },\quad \mu _{1,2,3}=1\dots 4;\;\mu _{1}<\mu
_{2}<\mu
_{3}\,,
\]
giving all in all 12 traceless and symmetric linear combinations. It
turns out that some of the operators in Eqns.~(\ref{eq:Opn3}) mix with
different non-symmetric representations
(cf.~Ref.~\cite{Gockeler:1996mu}) under renormalization.  Fortunately,
the mixing coefficient turns out to be negligible, at least for the
specific case considered in~\cite{Gockeler:1996wg}. It has to be
checked if this holds true for all potential mixing candidates in the
$n=3$ case.

The choice of index combinations for the construction of the H(4)
operators is not unique.  However, in the continuum limit all options
will lead to the same GFFs.  The interesting question of
determining an optimal set of operators that minimizes lattice
artifacts is beyond the scope of this presentation.

\subsection{\label{sec:lattice-momenta}Lattice momenta}
The possible values of the nucleon momenta $P$ and $P^{\prime }$ and
therefore the momentum transfer squared $t$ are severely restricted by the
lattice momenta.  The general three-momentum on a periodic lattice is
given by
\[
\overrightarrow{P}=\frac{2\pi }{aL_{s}}\overrightarrow{p}=\frac{2\pi
}{aL_{s}%
}\left( p_{1},p_{2},p_{3}\right) ,\quad
p_{i}=-L_{s},-L_{s}+1,\dots,L_{s}-1,L_{s}\,,
\]
where $a$ is the lattice spacing in GeV$^{-1}$ (we set $a=1$ in all
intermediate steps), and $L_{s}$ is the spatial lattice extension,
which is in our case $16$. It is clear that the lattice spacing times
the spatial extension determines the smallest possible non-zero
momentum, while the lattice spacing alone fixes the largest available
momentum. In practice, however, we are even more restricted to values
$p_{i}\ll L_{s}$, because large momenta lead to considerable noise.
For the nucleon four momenta $P$ and $P^{\prime }$, we use the
continuum dispersion relation, $P_{0}=P_{4}=%
\sqrt{m^{2}+\overrightarrow{P}^{2}}$, where $m$ is the nucleon mass,
determined in our lattice calculation. In order to construct the
maximally determined set of linear equations, Eq.~(\ref{eq:gff-soe1}),
we need to know all $ \overrightarrow{p},\overrightarrow{p}^{\prime }$
leading to the same momentum transfer squared $t$. So far we have
propagators available for the two sink momenta
$\overrightarrow{p}^{\prime }=\left( 0,0,0\right) $ and $%
\overrightarrow{p}^{\prime }=\left( -1,0,0\right) $, and the value for
$ \overrightarrow{q}\equiv \overrightarrow{p}^{\prime
}-\overrightarrow{p}$ is restricted by $\overrightarrow{q}^{2}<6$. We
plan to extend the allowed values in future investigations. The total
number of different $t$ is then 16, including the forward case.

Note, that $t$ depends in a nontrivial manner on
$\overrightarrow{p}^{\prime }$ and $\overrightarrow{p}$ individually
through the energies $P_{4},P_{4}^{\prime }$, and not only on the
relative momentum $\overrightarrow{q}$. For 8 of the 16 possible
values of $t$, we encountered highly visible fluctuations in the
nucleon propagators with negative values for the two point functions
at large $\tau \gg \tau _{\text{src}}$.  We excluded the corresponding
momentum combinations from our analysis. A list of all remaining
$\overrightarrow{q},\overrightarrow{p}^{\prime },\overrightarrow{p}$
and $t$, is shown in table~\ref{tab:all-latt-moms}.
\begin{table}
  \caption{\label{tab:all-latt-moms}Lattice momenta used in our
    calculation.}
  \begin{ruledtabular}
    \begin{tabular}{*{4}{l}}
      $\overrightarrow{p}^{\prime }$ & $\overrightarrow{q}$ & $\overrightarrow{p}$
      & $t$ [GeV$^2$] \\ \hline
      $
      \begin{array}{l}
        \left( 0,0,0\right)  \\ 
        \left( -1,0,0\right) 
      \end{array}
      $ & $
      \begin{array}{l}
        \left( 0,0,0\right)  \\ 
        \left( 0,0,0\right) 
      \end{array}
      $ & $
      \begin{array}{l}
        \left( 0,0,0\right)  \\ 
        \left( -1,0,0\right) 
      \end{array}
      $ & 0 \\ \hline
      $
      \begin{array}{l}
        \left( 0,0,0\right)  \\ 
        \left( -1,0,0\right) 
      \end{array}
      $ & $
      \begin{array}{l}
        \left( \pm 1,0,0\right) ,\left( 0,\pm 1,0\right) ,\left( 0,0,\pm 1\right) 
        \\ 
        \left( -1,0,0\right) 
      \end{array}
      $ & $
      \begin{array}{l}
        \left( \mp 1,0,0\right) ,\left( 0,\mp 1,0\right) ,\left( 0,0,\mp 1\right) 
        \\ 
        \left( 0,0,0\right) 
      \end{array}
      $ & -0.592 \\ \hline
      $\left( -1,0,0\right) $ & $\left( 0,\pm 1,0\right) ,\left( 0,0,\pm 1\right) $
      & $\left( -1,\mp 1,0\right) ,\left( -1,0,\mp 1\right) $ & -0.597 \\ 
      \hline
      $\left( 0,0,0\right) $ & $\left( \pm 1,\pm 1,0\right) ,\left( \pm 1,0,\pm
        1\right) ,\left( 0,\pm 1,\pm 1\right) $ & $\left( \mp 1,\mp 1,0\right)
      ,\left( \mp 1,0,\mp 1\right) ,\left( 0,\mp 1,\mp 1\right) $ & -1.134 \\ 
      \hline
      $\left( -1,0,0\right) $ & $\left( -1,\pm 1,0\right) ,\left( -1,0,\pm
        1\right) $ & $\left( 0,\mp 1,0\right) ,\left( 0,0,\mp 1\right) $ & -1.246 \\ 
      \hline
      $\left( -1,0,0\right) $ & $\left( -1,\pm 1,\pm 1\right) $ & $\left( 0,\mp
        1,\mp 1\right) $ & -1.844 \\ \hline
      $\left( -1,0,0\right) $ & $\left( -2,0,0\right) $ & $\left( 1,0,0\right) $ & 
      -2.492 \\ \hline
      $\left( -1,0,0\right) $ & $\left( -2,\pm 1,0\right) ,\left( -2,0,\pm
        1\right) $ & $\left( 1,\mp 1,0\right) ,\left( 1,0,\mp 1\right) $ & -3.090 \\ 
    \end{tabular}
  \end{ruledtabular}
\end{table}

%
%

\bibliography{GPD_moments}

\begin{thebibliography}{30}
\expandafter\ifx\csname natexlab\endcsname\relax\def\natexlab#1{#1}\fi
\expandafter\ifx\csname bibnamefont\endcsname\relax
  \def\bibnamefont#1{#1}\fi
\expandafter\ifx\csname bibfnamefont\endcsname\relax
  \def\bibfnamefont#1{#1}\fi
\expandafter\ifx\csname citenamefont\endcsname\relax
  \def\citenamefont#1{#1}\fi
\expandafter\ifx\csname url\endcsname\relax
  \def\url#1{\texttt{#1}}\fi
\expandafter\ifx\csname urlprefix\endcsname\relax\def\urlprefix{URL }\fi
\providecommand{\bibinfo}[2]{#2}
\providecommand{\eprint}[2][]{\url{#2}}

\bibitem[{\citenamefont{M{\"u}ller et~al.}(1994)\citenamefont{M{\"u}ller,
  Robaschik, Geyer, Dittes, and Horejsi}}]{Muller:1994fv}
\bibinfo{author}{\bibfnamefont{D.}~\bibnamefont{M{\"u}ller}},
  \bibinfo{author}{\bibfnamefont{D.}~\bibnamefont{Robaschik}},
  \bibinfo{author}{\bibfnamefont{B.}~\bibnamefont{Geyer}},
  \bibinfo{author}{\bibfnamefont{F.~M.} \bibnamefont{Dittes}},
  \bibnamefont{and} \bibinfo{author}{\bibfnamefont{J.}~\bibnamefont{Horejsi}},
  \bibinfo{journal}{Fortschr. Phys.} \textbf{\bibinfo{volume}{42}},
  \bibinfo{pages}{101} (\bibinfo{year}{1994}),
\eprint{hep-ph/9812448}.

\bibitem[{\citenamefont{Ji}(1997)}]{Ji:1997ek}
\bibinfo{author}{\bibfnamefont{X.-D.} \bibnamefont{Ji}},
  \bibinfo{journal}{Phys. Rev. Lett.} \textbf{\bibinfo{volume}{78}},
  \bibinfo{pages}{610} (\bibinfo{year}{1997}),
\eprint{hep-ph/9603249}.

\bibitem[{\citenamefont{Radyushkin}(1997)}]{Radyushkin:1997ki}
\bibinfo{author}{\bibfnamefont{A.~V.} \bibnamefont{Radyushkin}},
  \bibinfo{journal}{Phys. Rev.} \textbf{\bibinfo{volume}{D56}},
  \bibinfo{pages}{5524} (\bibinfo{year}{1997}),
\eprint{hep-ph/9704207}.

\bibitem[{\citenamefont{Burkardt}(2000)}]{Burkardt:2000za}
\bibinfo{author}{\bibfnamefont{M.}~\bibnamefont{Burkardt}},
  \bibinfo{journal}{Phys. Rev.} \textbf{\bibinfo{volume}{D62}},
  \bibinfo{pages}{071503} (\bibinfo{year}{2000}),
\eprint{hep-ph/0005108}.

\bibitem[{\citenamefont{Ralston and Pire}(2002)}]{Ralston:2001xs}
\bibinfo{author}{\bibfnamefont{J.~P.} \bibnamefont{Ralston}} \bibnamefont{and}
  \bibinfo{author}{\bibfnamefont{B.}~\bibnamefont{Pire}},
  \bibinfo{journal}{Phys. Rev.} \textbf{\bibinfo{volume}{D66}},
  \bibinfo{pages}{111501} (\bibinfo{year}{2002}),
\eprint{hep-ph/0110075}.

\bibitem[{\citenamefont{Diehl}(2002)}]{Diehl:2002he}
\bibinfo{author}{\bibfnamefont{M.}~\bibnamefont{Diehl}}, \bibinfo{journal}{Eur.
  Phys. J.} \textbf{\bibinfo{volume}{C25}}, \bibinfo{pages}{223}
  (\bibinfo{year}{2002}),
\eprint{hep-ph/0205208}.

\bibitem[{\citenamefont{Lai et~al.}(1997)}]{Lai:1997mg}
\bibinfo{author}{\bibfnamefont{H.~L.} \bibnamefont{Lai}} \bibnamefont{et~al.},
  \bibinfo{journal}{Phys. Rev.} \textbf{\bibinfo{volume}{D55}},
  \bibinfo{pages}{1280} (\bibinfo{year}{1997}),
\eprint{hep-ph/9606399}.

\bibitem[{\citenamefont{Gl{\"u}ck et~al.}(1998)\citenamefont{Gl{\"u}ck, Reya,
  and Vogt}}]{Gluck:1998xa}
\bibinfo{author}{\bibfnamefont{M.}~\bibnamefont{Gl{\"u}ck}},
  \bibinfo{author}{\bibfnamefont{E.}~\bibnamefont{Reya}}, \bibnamefont{and}
  \bibinfo{author}{\bibfnamefont{A.}~\bibnamefont{Vogt}},
  \bibinfo{journal}{Eur. Phys. J.} \textbf{\bibinfo{volume}{C5}},
  \bibinfo{pages}{461} (\bibinfo{year}{1998}),
\eprint{hep-ph/9806404}.

\bibitem[{\citenamefont{Martin et~al.}(2002)\citenamefont{Martin, Roberts,
  Stirling, and Thorne}}]{Martin:2001es}
\bibinfo{author}{\bibfnamefont{A.~D.} \bibnamefont{Martin}},
  \bibinfo{author}{\bibfnamefont{R.~G.} \bibnamefont{Roberts}},
  \bibinfo{author}{\bibfnamefont{W.~J.} \bibnamefont{Stirling}},
  \bibnamefont{and} \bibinfo{author}{\bibfnamefont{R.~S.}
  \bibnamefont{Thorne}}, \bibinfo{journal}{Eur. Phys. J.}
  \textbf{\bibinfo{volume}{C23}}, \bibinfo{pages}{73} (\bibinfo{year}{2002}),
\eprint{hep-ph/0110215}.

\bibitem[{\citenamefont{G{\"o}ckeler
  et~al.}(1996{\natexlab{a}})}]{Gockeler:1996wg}
\bibinfo{author}{\bibfnamefont{M.}~\bibnamefont{G{\"o}ckeler}}
  \bibnamefont{et~al.}, \bibinfo{journal}{Phys. Rev.}
  \textbf{\bibinfo{volume}{D53}}, \bibinfo{pages}{2317}
  (\bibinfo{year}{1996}{\natexlab{a}}),
\eprint{hep-lat/9508004}.

\bibitem[{\citenamefont{Best et~al.}(1997)}]{Best:1997ab}
\bibinfo{author}{\bibfnamefont{C.}~\bibnamefont{Best}} \bibnamefont{et~al.}
  (\bibinfo{year}{1997}),
\eprint{hep-ph/9706502}.

\bibitem[{\citenamefont{G{\"o}ckeler et~al.}(1997)}]{Gockeler:1997jk}
\bibinfo{author}{\bibfnamefont{M.}~\bibnamefont{G{\"o}ckeler}}
  \bibnamefont{et~al.}, \bibinfo{journal}{Nucl. Phys. Proc. Suppl.}
  \textbf{\bibinfo{volume}{53}}, \bibinfo{pages}{81} (\bibinfo{year}{1997}),
\eprint{hep-lat/9608046}.

\bibitem[{\citenamefont{G{\"o}ckeler et~al.}(2001)\citenamefont{G{\"o}ckeler,
  Horsley, Pleiter, Rakow, and Schierholz}}]{Gockeler:2001us}
\bibinfo{author}{\bibfnamefont{M.}~\bibnamefont{G{\"o}ckeler}},
  \bibinfo{author}{\bibfnamefont{R.}~\bibnamefont{Horsley}},
  \bibinfo{author}{\bibfnamefont{D.}~\bibnamefont{Pleiter}},
  \bibinfo{author}{\bibfnamefont{P.~E.~L.} \bibnamefont{Rakow}},
  \bibnamefont{and}
  \bibinfo{author}{\bibfnamefont{G.}~\bibnamefont{Schierholz}}
  (\bibinfo{year}{2001}),
\eprint{hep-ph/0108105}.

\bibitem[{\citenamefont{Dolgov et~al.}(2002)}]{Dolgov:2002zm}
\bibinfo{author}{\bibfnamefont{D.}~\bibnamefont{Dolgov}} \bibnamefont{et~al.}
  (\bibinfo{collaboration}{LHPC}), \bibinfo{journal}{Phys. Rev.}
  \textbf{\bibinfo{volume}{D66}}, \bibinfo{pages}{034506}
  (\bibinfo{year}{2002}),
\eprint{hep-lat/0201021}.

\bibitem[{\citenamefont{Detmold et~al.}(2001)\citenamefont{Detmold,
  Melnitchouk, Negele, Renner, and Thomas}}]{Detmold:2001jb}
\bibinfo{author}{\bibfnamefont{W.}~\bibnamefont{Detmold}},
  \bibinfo{author}{\bibfnamefont{W.}~\bibnamefont{Melnitchouk}},
  \bibinfo{author}{\bibfnamefont{J.~W.} \bibnamefont{Negele}},
  \bibinfo{author}{\bibfnamefont{D.~B.} \bibnamefont{Renner}},
  \bibnamefont{and} \bibinfo{author}{\bibfnamefont{A.~W.}
  \bibnamefont{Thomas}}, \bibinfo{journal}{Phys. Rev. Lett.}
  \textbf{\bibinfo{volume}{87}}, \bibinfo{pages}{172001}
  (\bibinfo{year}{2001}),
\eprint{hep-lat/0103006}.

\bibitem[{\citenamefont{Belitsky et~al.}(2002)\citenamefont{Belitsky,
  M{\"u}ller, and Kirchner}}]{Belitsky:2001ns}
\bibinfo{author}{\bibfnamefont{A.~V.} \bibnamefont{Belitsky}},
  \bibinfo{author}{\bibfnamefont{D.}~\bibnamefont{M{\"u}ller}},
  \bibnamefont{and} \bibinfo{author}{\bibfnamefont{A.}~\bibnamefont{Kirchner}},
  \bibinfo{journal}{Nucl. Phys.} \textbf{\bibinfo{volume}{B629}},
  \bibinfo{pages}{323} (\bibinfo{year}{2002}),
\eprint{hep-ph/0112108}.

\bibitem[{\citenamefont{Teukolsky et~al.}(1992)\citenamefont{Teukolsky,
  Vetterling, and Flannery}}]{Teukolsky:1992nr}
\bibinfo{author}{\bibfnamefont{S.~A.} \bibnamefont{Teukolsky}},
  \bibinfo{author}{\bibfnamefont{W.~T.} \bibnamefont{Vetterling}},
  \bibnamefont{and} \bibinfo{author}{\bibfnamefont{B.~P.}
  \bibnamefont{Flannery}}, \emph{\bibinfo{title}{Numerical Recipes in C}}
  (\bibinfo{publisher}{Cambridge University Press}, \bibinfo{year}{1992}),
  \bibinfo{edition}{2nd} ed.

\bibitem[{\citenamefont{Schroers}(2001)}]{Schroers:2001ph}
\bibinfo{author}{\bibfnamefont{W.}~\bibnamefont{Schroers}},
  \bibinfo{type}{Dissertation}, \bibinfo{school}{Fachbereich Physik, Bergische
  Uni\-ver\-si\-t{\"a}t-Ge\-samt\-hoch\-schule Wuppertal}
  (\bibinfo{year}{2001}),
\eprint{hep-lat/0304016}.

\bibitem[{\citenamefont{Schroers and {QCDSF
  collaboration}}(2002)}]{Schroers:2002ta}
\bibinfo{author}{\bibfnamefont{W.}~\bibnamefont{Schroers}} \bibnamefont{and}
  \bibinfo{author}{\bibnamefont{{QCDSF collaboration}}} (\bibinfo{year}{2002}),
  \bibinfo{note}{presented at the Workshop on ``Hadronic phenomenology from
  lattice gauge theory'' at the University of Regensburg, August 1-3, 2002.}

\bibitem[{\citenamefont{G{\"o}ckeler
  et~al.}(2003{\natexlab{a}})}]{Gockeler:2003ay}
\bibinfo{author}{\bibfnamefont{M.}~\bibnamefont{G{\"o}ckeler}}
  \bibnamefont{et~al.} (\bibinfo{collaboration}{QCDSF})
  (\bibinfo{year}{2003}{\natexlab{a}}),
\eprint{hep-lat/0303019}.

\bibitem[{\citenamefont{Kivel et~al.}(2001)\citenamefont{Kivel, Polyakov, and
  Vanderhaeghen}}]{Kivel:2000fg}
\bibinfo{author}{\bibfnamefont{N.}~\bibnamefont{Kivel}},
  \bibinfo{author}{\bibfnamefont{M.~V.} \bibnamefont{Polyakov}},
  \bibnamefont{and}
  \bibinfo{author}{\bibfnamefont{M.}~\bibnamefont{Vanderhaeghen}},
  \bibinfo{journal}{Phys. Rev.} \textbf{\bibinfo{volume}{D63}},
  \bibinfo{pages}{114014} (\bibinfo{year}{2001}),
\eprint{hep-ph/0012136}.

\bibitem[{\citenamefont{Brodsky et~al.}(2001)\citenamefont{Brodsky, Hwang, Ma,
  and Schmidt}}]{Brodsky:2000ii}
\bibinfo{author}{\bibfnamefont{S.~J.} \bibnamefont{Brodsky}},
  \bibinfo{author}{\bibfnamefont{D.~S.} \bibnamefont{Hwang}},
  \bibinfo{author}{\bibfnamefont{B.-Q.} \bibnamefont{Ma}}, \bibnamefont{and}
  \bibinfo{author}{\bibfnamefont{I.}~\bibnamefont{Schmidt}},
  \bibinfo{journal}{Nucl. Phys.} \textbf{\bibinfo{volume}{B593}},
  \bibinfo{pages}{311} (\bibinfo{year}{2001}),
\eprint{hep-th/0003082}.

\bibitem[{\citenamefont{Teryaev}(1998)}]{Teryaev:1998iw}
\bibinfo{author}{\bibfnamefont{O.~V.} \bibnamefont{Teryaev}}
  (\bibinfo{year}{1998}),
\eprint{hep-ph/9803403}.

\bibitem[{\citenamefont{Mathur et~al.}(2000)\citenamefont{Mathur, Dong, Liu,
  Mankiewicz, and Mukhopadhyay}}]{Mathur:1999uf}
\bibinfo{author}{\bibfnamefont{N.}~\bibnamefont{Mathur}},
  \bibinfo{author}{\bibfnamefont{S.~J.} \bibnamefont{Dong}},
  \bibinfo{author}{\bibfnamefont{K.~F.} \bibnamefont{Liu}},
  \bibinfo{author}{\bibfnamefont{L.}~\bibnamefont{Mankiewicz}},
  \bibnamefont{and} \bibinfo{author}{\bibfnamefont{N.~C.}
  \bibnamefont{Mukhopadhyay}}, \bibinfo{journal}{Phys. Rev.}
  \textbf{\bibinfo{volume}{D62}}, \bibinfo{pages}{114504}
  (\bibinfo{year}{2000}),
\eprint{hep-ph/9912289}.

\bibitem[{\citenamefont{G{\"o}ckeler
  et~al.}(2003{\natexlab{b}})\citenamefont{G{\"o}ckeler, Horsley, Pleiter,
  Rakow, Sch{\"a}fer, Schierholz, and Schroers}}]{Gockeler:2003ar}
\bibinfo{author}{\bibfnamefont{M.}~\bibnamefont{G{\"o}ckeler}},
  \bibinfo{author}{\bibfnamefont{R.}~\bibnamefont{Horsley}},
  \bibinfo{author}{\bibfnamefont{D.}~\bibnamefont{Pleiter}},
  \bibinfo{author}{\bibfnamefont{P.~E.} \bibnamefont{Rakow}},
  \bibinfo{author}{\bibfnamefont{A.}~\bibnamefont{Sch{\"a}fer}},
  \bibinfo{author}{\bibfnamefont{G.}~\bibnamefont{Schierholz}},
  \bibnamefont{and} \bibinfo{author}{\bibfnamefont{W.}~\bibnamefont{Schroers}}
  (\bibinfo{year}{2003}{\natexlab{b}}),
\eprint{hep-ph/0304249}.

\bibitem[{\citenamefont{Airapetian et~al.}(2001)}]{Airapetian:2001yk}
\bibinfo{author}{\bibfnamefont{A.}~\bibnamefont{Airapetian}}
  \bibnamefont{et~al.} (\bibinfo{collaboration}{HERMES}),
  \bibinfo{journal}{Phys. Rev. Lett.} \textbf{\bibinfo{volume}{87}},
  \bibinfo{pages}{182001} (\bibinfo{year}{2001}),
\eprint{hep-ex/0106068}.

\bibitem[{\citenamefont{Stepanyan et~al.}(2001)}]{Stepanyan:2001sm}
\bibinfo{author}{\bibfnamefont{S.}~\bibnamefont{Stepanyan}}
  \bibnamefont{et~al.} (\bibinfo{collaboration}{CLAS}), \bibinfo{journal}{Phys.
  Rev. Lett.} \textbf{\bibinfo{volume}{87}}, \bibinfo{pages}{182002}
  (\bibinfo{year}{2001}),
\eprint{hep-ex/0107043}.

\bibitem[{\citenamefont{Favart}(2002)}]{Favart:2002eh}
\bibinfo{author}{\bibfnamefont{L.}~\bibnamefont{Favart}},
  \bibinfo{journal}{Nucl. Phys.} \textbf{\bibinfo{volume}{A711}},
  \bibinfo{pages}{165} (\bibinfo{year}{2002}),
\eprint{hep-ex/0207030}.

\bibitem[{\citenamefont{Montvay and M{\"u}nster}(1994)}]{Montvay:1994cy}
\bibinfo{author}{\bibfnamefont{I.}~\bibnamefont{Montvay}} \bibnamefont{and}
  \bibinfo{author}{\bibfnamefont{G.}~\bibnamefont{M{\"u}nster}},
  \emph{\bibinfo{title}{Quantum Fields on a Lattice}}, Cambridge Monographs on
  Mathematical Physics (\bibinfo{publisher}{Cambridge University Press},
  \bibinfo{year}{1994}).

\bibitem[{\citenamefont{G{\"o}ckeler
  et~al.}(1996{\natexlab{b}})}]{Gockeler:1996mu}
\bibinfo{author}{\bibfnamefont{M.}~\bibnamefont{G{\"o}ckeler}}
  \bibnamefont{et~al.}, \bibinfo{journal}{Phys. Rev.}
  \textbf{\bibinfo{volume}{D54}}, \bibinfo{pages}{5705}
  (\bibinfo{year}{1996}{\natexlab{b}}),
\eprint{hep-lat/9602029}.

\end{thebibliography}

%
%

\end{document}